\documentclass[10pt,conference]{IEEEtran}
\usepackage{cite}
\usepackage{amsmath,amssymb,amsfonts}
\usepackage{algorithmic}
\usepackage{graphicx}
\usepackage{textcomp}
\usepackage{xcolor}
\usepackage[hyphens]{url}
\usepackage{fancyhdr}
\usepackage{hyperref}
\usepackage{cite}
\usepackage{amsmath,amssymb,amsfonts}
\usepackage{algorithmic}
\usepackage{graphicx}
\usepackage{textcomp}
\usepackage{xcolor}
\usepackage{booktabs}
\usepackage{multirow}

\newcommand{\hpcayear}{2026}
\newcommand{\chingyi}[1]{\textcolor{black}{#1}}

\newcommand{\hpcasubmissionnumber}{1939}

\begin{document}
\title{Systolic Array-based Architecture for Low-Bit Integerized Vision Transformers}

\IEEEpeerreviewmaketitle

\def\hpcacameraready{} 
\newcommand{\hpcapubid}{0000--0000/00\$00.00}
\newcommand\hpcaauthors{Ching-Yi Lin$\dagger$ and Sahil Shah$\dagger$}
\newcommand\hpcaaffiliation{$\dagger$University of Maryland}
\newcommand\hpcaemail{\{chingyil, sshah389\}@umd.edu}



\author{
  \ifdefined\hpcacameraready
    \IEEEauthorblockN{\hpcaauthors{}}
      \IEEEauthorblockA{
        \hpcaaffiliation{} \\
        \hpcaemail{}
      }
  \else
    \IEEEauthorblockN{\normalsize{HPCA \hpcayear{} Submission
      \textbf{\#\hpcasubmissionnumber{}}} \\
      \IEEEauthorblockA{
        Confidential Draft \\
        Do NOT Distribute!!
      }
    }
  \fi 
}

\fancypagestyle{camerareadyfirstpage}{%
  \fancyhead{}
  \renewcommand{\headrulewidth}{0pt}
  \fancyhead[C]{
    \ifdefined\aeopen
    \parbox[][12mm][t]{13.5cm}{\hpcayear{} IEEE International Symposium on High-Performance Computer Architecture (HPCA)}    
    \else
      \ifdefined\aereviewed
      \parbox[][12mm][t]{13.5cm}{\hpcayear{} IEEE International Symposium on High-Performance Computer Architecture (HPCA)}
      \else
      \ifdefined\aereproduced
      \parbox[][12mm][t]{13.5cm}{\hpcayear{} IEEE International Symposium on High-Performance Computer Architecture (HPCA)}
      \else
      \parbox[][0mm][t]{13.5cm}{}
    \fi 
    \fi 
    \fi 
    \ifdefined\aeopen 
      \includegraphics[width=12mm,height=12mm]{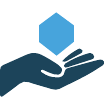}
    \fi 
    \ifdefined\aereviewed
      \includegraphics[width=12mm,height=12mm]{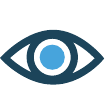}
    \fi 
    \ifdefined\aereproduced
      \includegraphics[width=12mm,height=12mm]{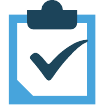}
    \fi
  }
  \fancyfoot[C]{}
}
\fancyhead{}
\renewcommand{\headrulewidth}{0pt}

\maketitle

\ifdefined\hpcacameraready 
  \thispagestyle{camerareadyfirstpage}
  \pagestyle{empty}
\else
  \thispagestyle{plain}
  \pagestyle{plain}
\fi

\newcommand{\hpcaheight}{0mm}
\ifdefined\eaopen
\renewcommand{\hpcaheight}{12mm}
\fi


\begin{abstract}
Transformer-based models are becoming more and more intelligent and are revolutionizing a wide range of human tasks. To support their deployment, AI labs offer inference services that consume hundreds of GWh of energy annually and charge users based on the number of tokens processed. Under this cost model, minimizing power consumption and maximizing throughput have become key design goals for the inference hardware. While graphics processing units (GPUs) are commonly used, their flexibility comes at the cost of low operational intensity and limited efficiency, especially under the high query-per-model ratios of modern inference services.

In this work, we address these challenges by proposing a low-bit, model-specialized accelerator that strategically selects tasks with high operation (OP) reuse and minimal communication overhead for offloading. Our design incorporates multiple systolic arrays with deep, fine-grained pipelines and array-compatible units that support essential operations in multi-head self-attention (MSA) module. At the accelerator-level, each self-attention (SA) head is pipelined within a single accelerator to increase data reuse and further minimize bandwidth.

Our 3-bit integerized model achieves 96.83\% accuracy on CIFAR-10 and 77.81\% top-1 accuracy on ImageNet. We validate the hardware design on a 16nm FPGA (Alveo U250), where it delivers 13,568 GigaOps/second (GOPs/s) and 219.4 GOPs/s/W. Compared to a same-technology GPU (GTX 1080), our design offers 1.50x higher throughput and 4.47x better power efficiency. Even against a state-of-the-art GPU (RTX 5090), we still achieve 20\% better power efficiency despite having 87\% lower throughput.
\end{abstract}


\section{Introduction}
Recent advancements in large language models (LLMs) have enabled them to achieve performance across a wide range of human tasks. To support their deployment, frontier AI laboratories offer inference services \chingyi{with hundreds of GWh of energy annually~\cite{Todorovic.2024}}.
\chingyi{For example, OpenAI reportedly spends \$4 billion on running inference workload alone~\cite{Moss.2024}. At the same time, such services typically charge users based on the number of tokens processed. Under this cost model, minimizing power consumption and maximizing throughput have become key design goals for the inference hardware. These two metrics are often combined into a single metric token/s/W in model performance evaluation~\cite{samsi2023words}.}

\chingyi{Currently, }GPUs are the predominant computing platforms employed by service providers. These hardware solutions offer high bandwidth and flexible programmability to accommodate a wide variety of machine learning models. However, \chingyi{this flexibility becomes less advantageous in LLM inference server due to modern huge query-per-model ratio over the lifetime of the model~\cite{sardana2023beyond}.}
In other words, with the vast number of inference requests for the same model, it is worthwhile to explore hardware specialized for a particular model for superior \chingyi{throughput and power efficiency}. 

\begin{figure}[!b]
    \centering
    \includegraphics[width=1.0\linewidth]{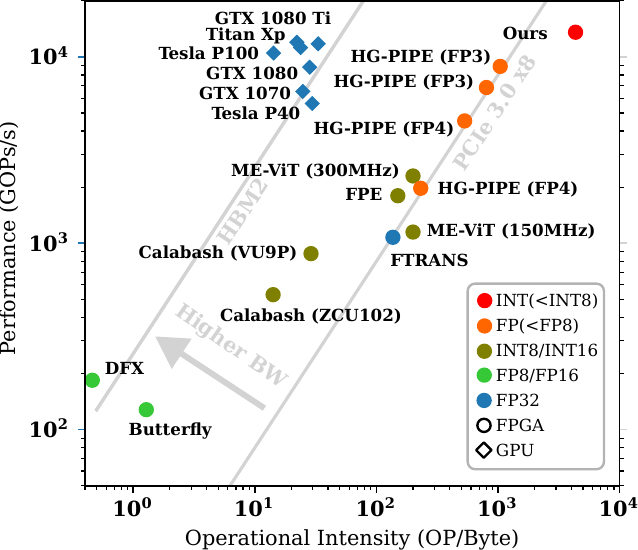}
    \caption{This figure evaluates the performance and bandwidth requirements of related works. Many of FPGA designs are restricted by communication speed due to their low operational intensity. GPUs achieve higher performance by maximizing bandwidth utilization. In this work, we improve operational intensity, allowing us to sustain high performance without increasing bandwidth demands.}
    \label{fig:roofline}
\end{figure}

Based on these considerations, the primary design goal for service-oriented hardware is to \chingyi{build a more model-customized architecture to leverage model-specific characteristics. Specifically, we focus on increasing} compute density and throughput within a low-bandwidth environment for a more scalable infrastructure. \chingyi{To meet these goals, model integerization has emerged as a promising strategy ~\cite{jacob2018quantization, lin2025low, li2023vit} to achieve both subgoals using lightweight integer arithmetic}. Prior researches have shown that \chingyi{convolutional} models can be quantized to 8-bit without observable accuracy loss ($<1\%$ top-1 accuracy on ImageNet)~\cite{jacob2018quantization} while still being compatible with mobile platforms. Nevertheless, \chingyi{their latency is not improved}
due to the limited improvement of their INT8 logic~\cite{jacob2018quantization}. In this work, we focus on low-bit networks \chingyi{($\leq$4-bit)}, which provide superior hardware performance
\chingyi{than 8-bit. While this comes with a slight reduction in model accuracy, t}his inherent tradeoff between accuracy and hardware cost aligns well with the diverse \chingyi{accuracy-cost options} offered by frontier AI labs. We specifically select 3-bit as our target precision, \chingyi{because prior work has demonstrated that a 3-bit quantized DeiT-S incurs} only a 1\% loss compared to the full-precision DeiT-S on the ImageNet dataset \cite{li2022q}. \chingyi{Following the integerization approach~\cite{lin2025low}, we build an integerized model where compute-intensive modules are realized in low-bit integer arithmetic}

Despite the clear benefits \chingyi{and successful low-bit integerization, the efficient inference of low-bit model} 
remains a challenge. \chingyi{To build a high-throughput, low-bandwidth, and power-efficient architecture, we analyze both FPGA- and GPU-based hardware built in the same technology (16nm.)}
\chingyi{As the roofline model~\cite{williams2009roofline} illustrated in Figure \ref{fig:roofline}, recent FPGA works align with PCIe 3.0 x8, which is the most common interface protocol for FPGAs. This alignment implies that the throughput of these designs is constrained by low operational intensity, which usually arises from the poor OP reuse in small-size operations to support a wide range of model.}
\chingyi{In this work, our design leverage model-specific properties of the target application}
to strategically select the tasks with \chingyi{maximized OP reuse and} low communication \chingyi{overhead} for offloading. Furthermore, the simpler logic from low-bit arithmetic results in fewer gates per multiply-accumulate (MAC)\chingyi{, leading to higher compute density and the ability to perform} more operations \chingyi{from a received packet.}

To achieve these aforementioned goals, we propose a \chingyi{systolic array-based} architecture with a deep and fine-grained pipeline \chingyi{that support} massive parallelism and high clock frequencies. \chingyi{The systolic array template includes support for aggregation operations such as layer normalization and softmax via array-compatible units,}
ensuring the template can be generalized to various models. \chingyi{To handle the cycle mismatch between arrays, we employ FIFOs to eliminate the need for cache or random memory access during the inference.}

Our main contributions are summarized as follows:
\begin{itemize}
    \item We propose a systolic array-based architecture for low-bit MSA, specifically optimized for operational intensity to achieve high throughput under the low bandwidth constraints.
    \item We design systolic array-compatible Softmax and Layer Normalization modules, allowing these aggregate operations to seamlessly integrate with our systolic array template.
    \item We optimize the power efficiency of the system through low-bit arithmetic and accelerator pipelining, surpassing state-of-the-art FPGA implementations.
    \item We conduct a detailed analysis for power and area, demonstrating that the \chingyi{MAC operations} primarily dominate power and area of this MSA accelerator.
\end{itemize}
\section{Background}
\subsection{Model Quantization}
Low storage and efficient arithmetic are two critical design objectives for ML model inference. Methods to reduce the model storage size generally fall into two categories. In the first category, often termed \textit{model compression}, aims to minimize model size but preserving the same accuracy. This method includes pruning the unused connection~\cite{luo2017thinet} or incorporating information entropy in loss term during training\cite{han2015deep}.

The other category focuses on training inherently low-information networks instead of reducing from high-precision models. Examples include Binarized Neural Network \cite{courbariaux2016binarized} and XNOR-Net \cite{rastegari2016xnor}, which constrain the model weights and activations to $\{-1,1\}$. While Ternary Neural Network \cite{alemdar2017ternary} expands the quantized space to $\{-1,0,1\}$, the accuracy of these models are low ($<$90\% in CIFAR-10) due to their small information space in the model. To enable more generic low-information networks, DoReFa-Net \cite{zhou2016dorefa} achieved arbitrary-bit weight and activation using a straight-through estimator (STE). Choi \cite{choi2018bridging} further improved accuracy by employing statistics-aware weight binning (SAWB) and parameterized clipping activation (PACT), achieving over $74.2\%$ ImageNet accuracy with ResNet50.

For transformer-based models, FQ-ViT \cite{lin2021fq} reduced DeiT series models to $4$-bit attention map and $8$-bit weights within $2$\% accuracy loss. Q-ViT further pushed model precision to 3-bit with similar accuracy using \textit{switchable scale} \cite{li2022q}. Despite the impressive accuracy and reduced storage space of these quantized models, they still require dequantized operands during the inference, which \chingyi{not only} introduces computational overhead\chingyi{, but also makes the computational-intensive modules, such as linear layer and matrix multiplication, operate in high precision~\cite{li2023vit}}.

To achieve both low storage and efficient arithmetic simultaneously, integerized model have emerged as a widely studied solution to leverage the maturity of integer-arithmetic.

Jacob \textit{el al.} \cite{jacob2018quantization} first quantized MobileNet to 8-bit and achieved only $1-2\%$ accuracy loss on ImageNet. For transformer-based model, I-BERT \cite{kim2021bert} adopts polynomial approximation for Softmax and GeLU functions. I-ViT~\cite{li2023vit} further enhanced the performance using shift-based Softmax and GeLU. Huang \cite{huang2023integer} applied shift-based exponential and second-order polynomial approximation for improved hardware performance. Despite these advancements in integerized models, a critical gap remains between the 8-bit model integerization and a low-bit, efficient implementation. \chingyi{In this work, we propose a low-bit quantized model following a low-bit integerization algorithm~\cite{lin2025low}. Our method builds on the success of low-bit quantization and model finetuning by training an integerized model from a quantized model checkpoint, enabling both high accuracy and efficient inference}

\subsection{Systolic Array}
Systolic arrays are renowned for their inherent parallelism and pipelined dataflow, making them ideal for achieving high throughput and low memory bandwidth requirements. However, their highly regular structure often limits their flexibility when adapting to the diverse compute graphs of various models.

The design of systolic arrays typically focuses on two optimization goals.
The first is to maximize the utilization rate of processing elements (PEs). While some efforts have increased utilization rate by parallelizing multiple operation within a single PE \cite{ye2023accelerating,huang2023integer}, the utilization rate remains bounded by the complicated and varied operations in convolutional or transformer networks. Building on this, Calabash~\cite{luo2023calabash} improved the utilization rate at the module-level by multiplexing systolic arrays, enabling different operations through reconfigurable configurations. At the PE-level, Zhang employed weight-loop computing unit (WLCU) with a bubble-free dataflow to maximize the utilization rate~\cite{zhang2024109}. In addition to customized flow, high-level synthesis (HLS) tools are also used to optimize the performance~\cite{li2022auto, ye2023accelerating}.

The second design goal is to increase the operational intensity by maximizing reuse and translate it into higher throughput from a given bandwidth~\cite{marino2023me}. For example, Eyeriss~\cite{chen2016eyeriss} proposed row-stationary dataflow and utilized local scratchpad memories to maintain the model parameter locally and reduce data transfers. Similarly, in transformer-based model, ME-ViT~\cite{marino2023me} also introduced a single-load policy to buffer weights and intermediate data to minimize external data traffic.

\chingyi{In this work, we focus on the second goal to optimize the operational intensity. We exploit the regular architecture from transformer models to select high-operational-intensity cut point. This minimizes off-chip traffic and enables us to achieve operational intensity higher than 4000 OPs/byte.}

\begin{figure}
    \centering
    \includegraphics[width=\linewidth]{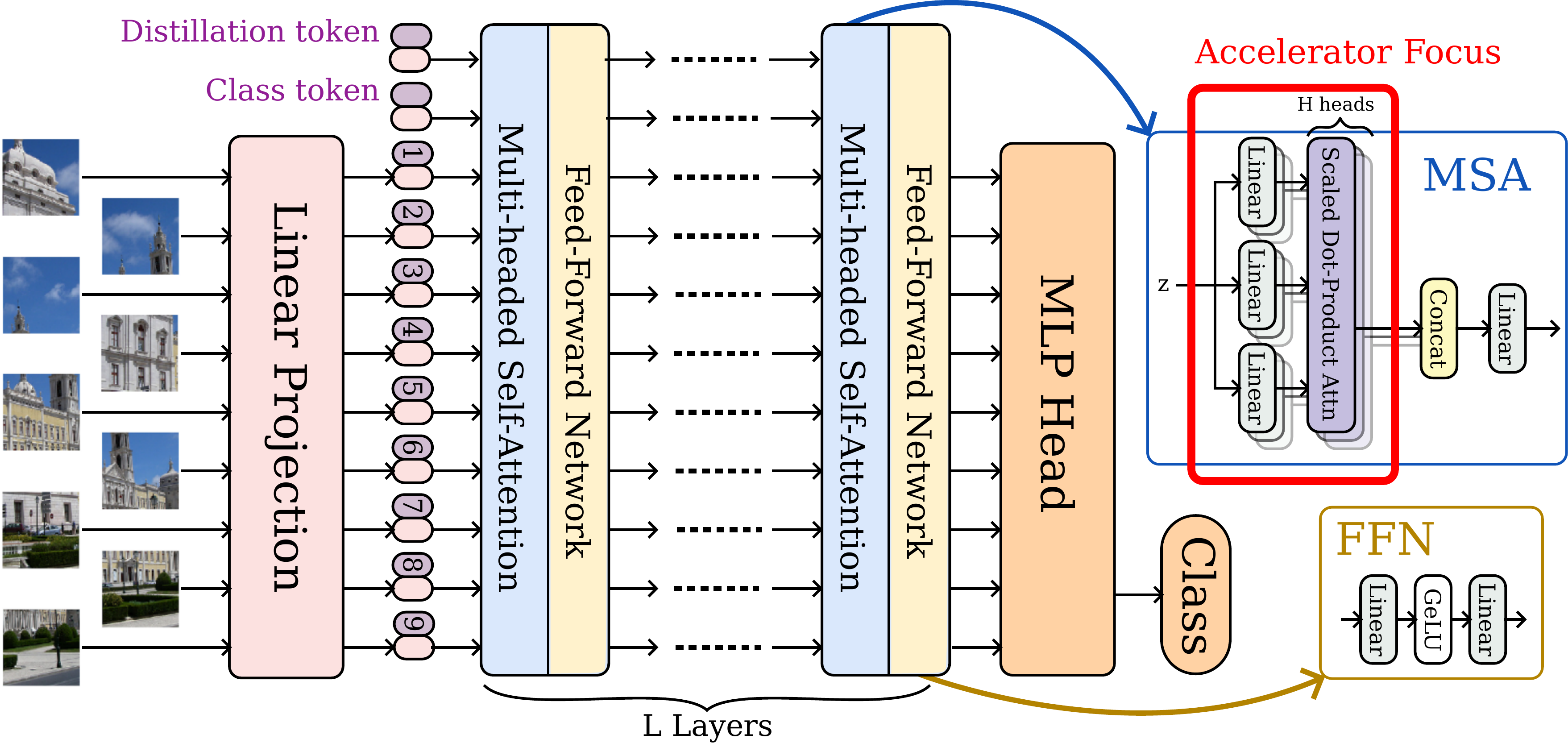}
    \caption{DeiT structure consists of linear projection of flattened patches, $L$-layer transformer layers, and a multi-layer perceptron (MLP) head. Each transformer layer has a MSA module and a FFN module. Our accelerator design focuses on the QKV projection and scaled dot-product attention in MSA.}
    \label{fig:deit-full}
\end{figure}

\section{Integerized Model}
\subsection{Visual Transformer (ViT) Overview}
DeiT model, the target model in this work, consists of MSA and feed-forward network (FFN), as illustrated in Figure \ref{fig:deit-full}. The difference between Figure \ref{fig:deit-full} and ViT~\cite{dosovitskiy2020image} is the extra distillation token at the input. Our accelerator design focuses on the MSA module, as it contains linear layer, matrix multiplication, Softmax, and layer normalization, covering all the operations in FFN except for an element-wise non-linear function GeLU. 

Built from single SA, the SA result can be obtained from input projection (Eq. \ref{eqn:qkv}), getting attention score (Eq. \ref{eqn:vit-attn}), and dot-product attention (Eq. \ref{eqn:vit-attn}).
\begin{equation} \label{eqn:qkv}
    [\mathbf{Q}, \mathbf{K}, \mathbf{V}] = \mathbf{zU}_{qkv}
\end{equation}
\begin{equation}
    A = \text{softmax} (\widetilde{\mathbf{Q}}\cdot\widetilde{\mathbf{K}}^T/D_h^{1/2})\text{, where }\widetilde{\cdot}=LN(\cdot)
\end{equation}
\begin{equation}
    SA(\mathbf{z}) = A\mathbf{V}
    \label{eqn:vit-attn}
\end{equation}
Here we follow the expression in ViT~\cite{dosovitskiy2020image} with $\cdot$ operator explicitly for matrix multiplication. The difference is the additional layer normalization in Equation \ref{eqn:vit-attn}. The post-normalized model has been shown to achieve high data efficiency in several applications~\cite{touvron2021training}. The results from multiple SA modules are concatenated to obtain MSA results as in Equation \ref{eqn:vit-msa}. 
\begin{equation}
    MSA(\textbf{z}) = [SA_1(\mathbf{z});SA_2(\mathbf{z});\dots;SA_k(\mathbf{z})]\cdot\mathbf{U}_{msa}
    \label{eqn:vit-msa}
\end{equation}
\begin{equation}
    \mathbf{z}'_l = MSA(\mathbf{z}_{l-1}) + \mathbf{z}_{l-1}
    \label{eqn:residual}
\end{equation}
To reduce the parameter size, Q-ViT approximates $\mathbf{Q}$, $\mathbf{K}$, $\mathbf{V}$, $A$, and $SA_i(\mathbf{z})$ in a low-bit expression~\cite{li2022q}. A low-bit dequantzied expression $\hat{\mathbf{z}}$ can be approximated \chingyi{as the product of low-bit integer $\mathbf{z}_{3b}$ and a channel-wise step size $\Delta_z$}. \chingyi{This} step size is shared across all data in the same channel. The dequantization can be obtained in a matrix multiplication form $\mathbf{z}_{3b}\cdot\mathbf{I}_{\Delta z}$, making input projection of the MSA be expressed as

\begin{equation}
    [\mathbf{Q}, \mathbf{K}, \mathbf{V}] = \hat{\mathbf{z}}\mathbf{\hat{U}}_{qkv} = (\mathbf{z}_{3b}\cdot\mathbf{I}_{\Delta z})(\mathbf{U}_{qkv,3b}\cdot\mathbf{I}_{\Delta Umsa})
\end{equation}

\begin{figure}
    \centering
    \includegraphics[width=0.9\linewidth]{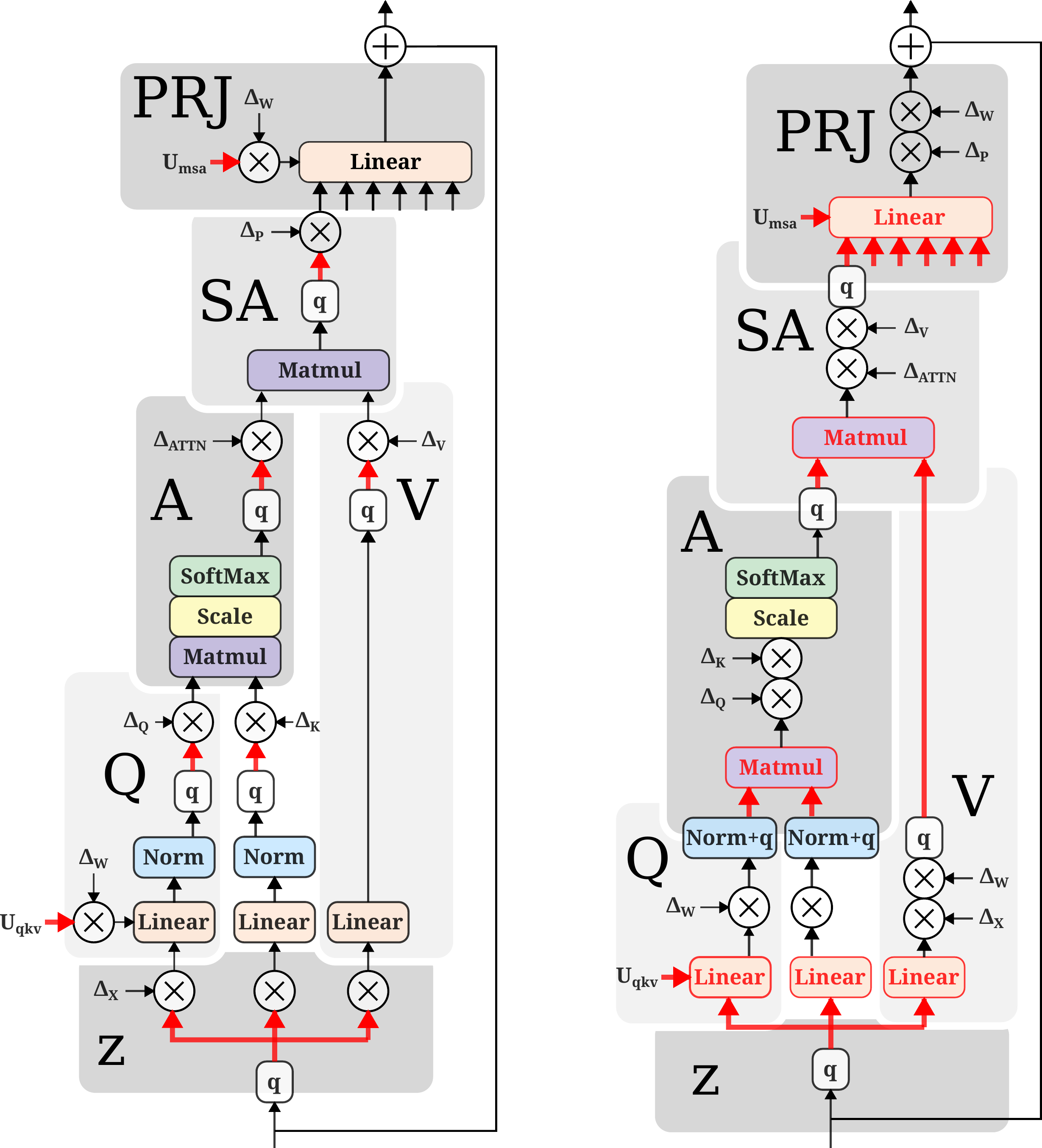}
    \caption{(A) The original compute graph of ViT shows that floating-point (FP) values are quantized into low-bit representations (in red), but the compute-intensive modules (Linear and MatMul) operate on dequantized FP values. (B) In the integerized model, the operations are reordered such that the compute-intensive modules receive low-bit inputs directly.}
    \label{fig:integerized-compute-graph}
\end{figure}

Although this quantization method can reduce the storage size, the operands of matrix multiplication $\mathbf{z}_{3b}\cdot\mathbf{I}_{\Delta z}$ and $\mathbf{U}_{qkv,3b}\cdot\mathbf{I}_{\Delta Umsa}$ are still floating point. This results in high computational use due to the square relationship between the number of gate and the bit~\cite{luo2016finegrained}.

To reduce the computational resource, we aim to factor out the step size and separate the matrix multiplication to a low-bit integerized matrix multiplication $\mathbf{z}_{3b}\cdot\mathbf{U}_{qkv,3b}$ with a post-dequantization. To achieve this, we reduce the channel-wise $\Delta_{z}$ to the global step size $\overline{\Delta z}$ to allow it to be factored out. The integerized matrix multiplication is expressed in Equation \ref{eqn:integer-vit}

\begin{align}
    [\mathbf{Q}, \mathbf{K}, \mathbf{V}] &= \hat{\mathbf{z}}\mathbf{\hat{U}}_{qkv}\\
    &= (\mathbf{z}_{3b}\cdot(\overline{\Delta z})\mathbf{I})\cdot(\mathbf{U}_{qkv,3b}\cdot\mathbf{I}_{\Delta Uqkv})\\
    &= (\mathbf{z}_{3b}\cdot\mathbf{U}_{qkv,3b})\cdot(\overline{\Delta z}\cdot\mathbf{I}_{\Delta Uqkv})\textbf{}
    \label{eqn:integer-vit}
\end{align}

The integerization technique can be applied to other operations, such as getting attention score and dot-product attention. Figure \ref{fig:integerized-compute-graph} shows the original compute graph and the integerized compute graph. To emphasize the effectiveness of the integerization, the low-bit data and the integer computation are highlighted in red. In the integerized compute graph, the compute-intensive modules, linear layer (Linear) and matrix multiplication (Matmul), are executed in low bit. Although normalization and softmax are still full precision, these modules have lower complexity ($O(N^2)$) than Linear and Matmul ($O(N^3)$).

\begin{figure}
    \centering
    \includegraphics[width=0.9\linewidth]{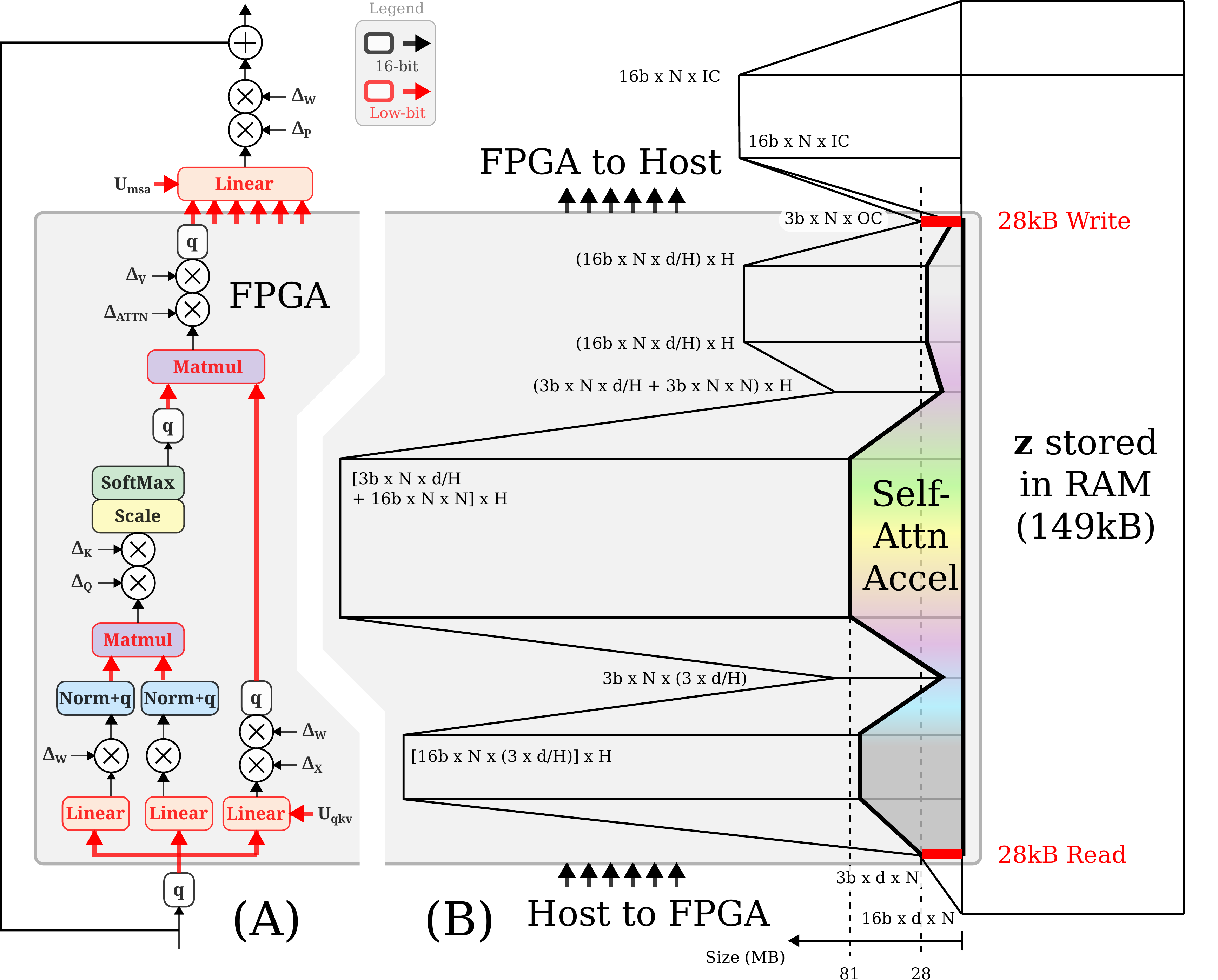}
    \caption{(A) To reduce communication overhead between the host and FPGA, the FPGA receives the quantized input $\mathbf{z}_{3b}$ and returns low-bit weighted sum results. (B) The communication load of a MSA module is analyzed to validate our strategy. The chosen cut point yields the smallest data size (28kB) among all intermediate results, and the full-precision $\mathbf{z}$ remain on the host until it is added to the projection output.}
    \label{fig:comm-load}
\end{figure}

\subsection{SA Offloading}

To maximize the operational intensity (OP/byte), the offloaded operation needs to (1) Minimize the size of the input/output (2) Maximize the data reuse. 

To maximize the data reuse, the operation needs to include the compute-intensive operations. In MSA, there are multiple SA modules and a linear module. Each SA includes two matrix multiplication and one linear module. As the ability to accommodate all these modules depends on the number of resources and device/technology dependent. We leave the problem until the implementation stage.

\chingyi{Figure \ref{fig:comm-load} shows the data size at each stage in SA to analyze their communication load. }$\mathbf{z}$ is stored until it is used in \chingyi{final addition (Equation \ref{eqn:residual})}, and the quantized copy is used to generate MSA result. The quantized $\mathbf{z}$, or $\mathbf{z}_{3b}$, is fed into FPGA due to its small size (28kB in DeiT-S). Similarly, the $SA_{3b}(\mathbf{z})$ is returned from the FPGA. One noting thing is, although the $MSA(\mathbf{z})$ is integerized to $([SA_{3b,1}(\mathbf{z});\cdot;SA_{3b,k}(\mathbf{z})]\cdot\mathbf{U}_{3b,msa})\cdot((\overline{\Delta_{SA}})\mathbf{I}_{\Delta Umsa})$, and the $SA_{3b}(\mathbf{z})$ is sent from the FPGA, other operations, including the concatenation, low-bit matrix multiplication, and dequantization, are all performed on the host side to save communication bandwidth.

\begin{figure}
    \centering
    \includegraphics[width=0.9\linewidth]{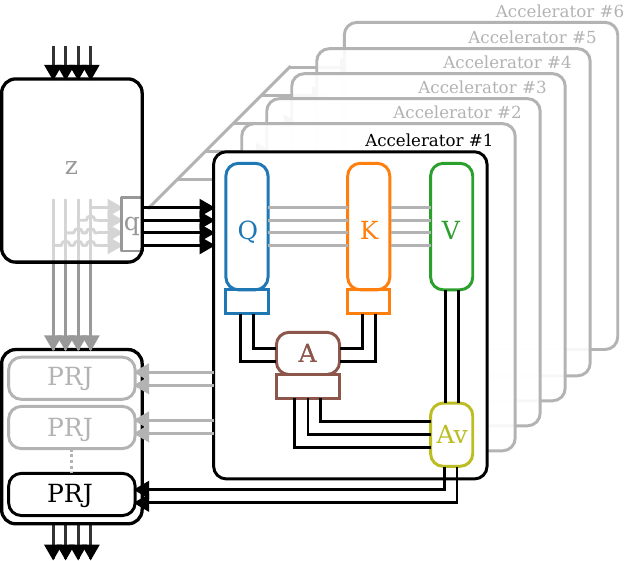}
    \caption{The quantized input $\mathbf{z}$ is offloaded to the accelerator, where the SA module is computed, including the generation of Q/K/V, attention scores with softmax ($\mathbf{A}$), and the weighted value product ($\mathbf{AV}$). The output of the SA module is then returned to the host, where projection ($\mathbf{PRJ}$) and aggregation are performed.}
    \label{fig:dataflow}
\end{figure}

\section{Accelerator Architecture}

\subsection{Dataflow Overview}
The high-level dataflow is illustrated in Figure \ref{fig:dataflow}. The accelerator receives input from a host-side quantizer, which converts the full-precision $\mathbf{z}$ into low-bit values $\mathbf{z}_{3b}$. These low-bit values are then sent to three systolic arrays to generate $\mathbf{Q}_{3b}$, $\mathbf{K}_{3b}$, and $\mathbf{V}_{3b}$ respectively. Some systolic array includes an additional module for aggregation operation such as normalization or softmax.

To reduce model-to-model communication overhead, all the high-precision data and intermediate results are both within individual modules. In the other words, only low-bit data are transmitted in the inter-module connection, or the black and gray lines in the accelerator in Figure \ref{fig:dataflow}. This reduces place-and-route efforts or simplifies further application-specific integrated circuit (ASIC) implementation.

\begin{figure}[b]
    \centering
    \includegraphics[width=0.9\linewidth]{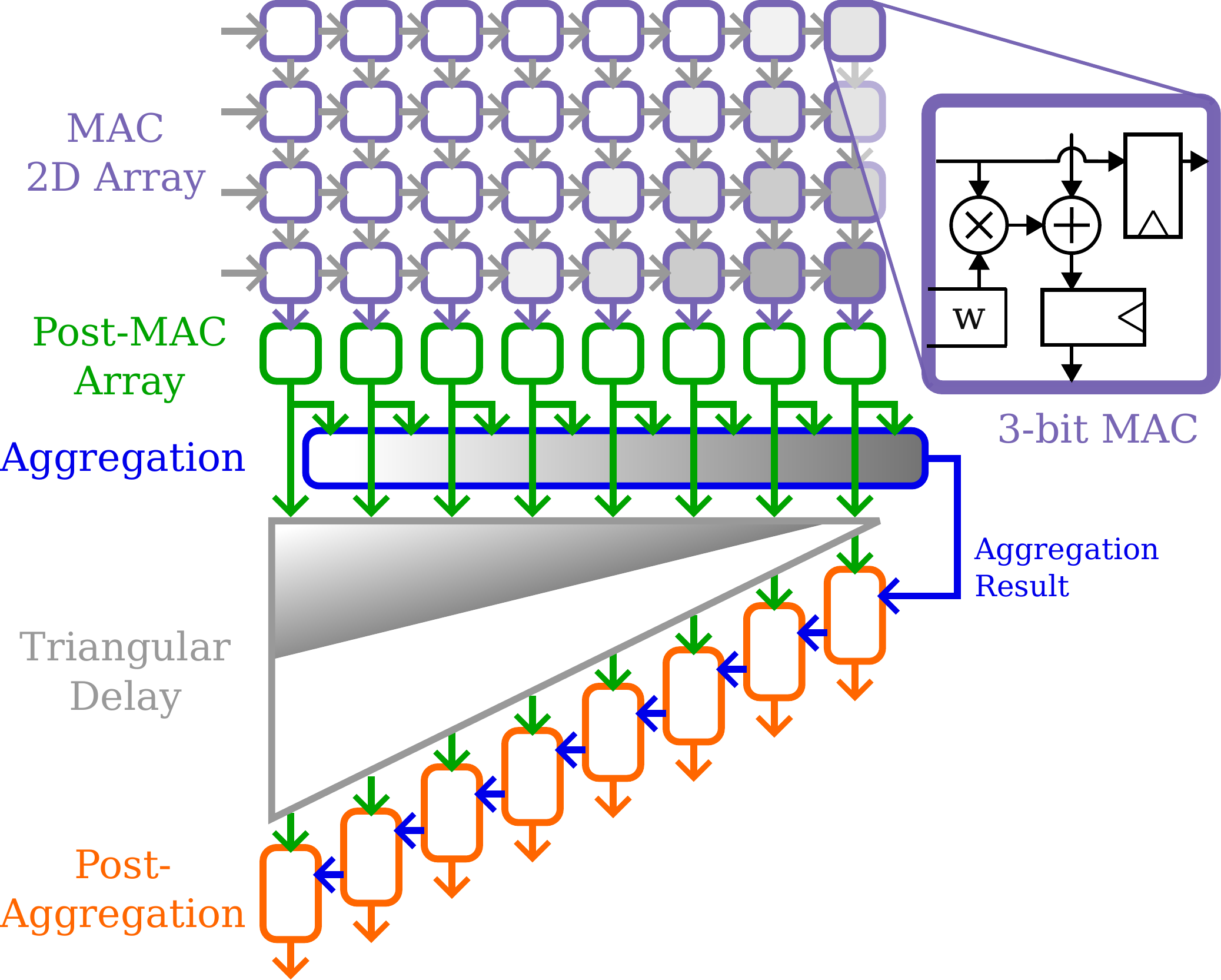}
    \caption{Systolic Array Overview: The template comprises a 2D MAC array, a post-MAC array, a systolic aggregation module, a triangular delay unit, and a post-aggregation array. The 2D MAC array is built from 3-bit MAC unit. The post-MAC array performs element-wise transformations on the output of the MAC array. The aggregation module then accumulates the post-MAC results using a systolic dataflow. This aggregated result is further propagated through the post-aggregation array, also in a systolic manner. A triangular delay is added to align the timing between the post-MAC and post-aggregation arrays.}
    \label{fig:array-overview}
\end{figure}

\subsection{Array Design}
\chingyi{Figure \ref{fig:array-overview} illustrates the architecture template. The array consists of a 2D MAC array, a post-MAC array, a systolic aggregation module, a triangular delay unit, and a post-aggregation array. The 2D MAC array performs the matrix multiplication $\mathbf{W}^T\mathbf{X}$ in a weight-stationary (WS) dataflow. The results are then processed by the post-MAC units. Next, the aggregation module performs row-wise aggregation on the post-MAC module output $f(\mathbf{W}^T\mathbf{X})$, and the aggregation results are propagated to post-aggregation module in a systolic manner. A major advantage of this architecture is its fully systolic nature: all inter-PE communication is local, short-range, and avoids costly broadcast transmission in data. This matches our aforementioned design goal to reduce the bandwidth requirement.}

\chingyi{Despite the simple template, this architecture is general enough to support all the operation in MSA, including scaling, quantization, layer normalization, and softmax. The following sections provide a detailed breakdown of each module.}

\subsubsection{Processing Elements (PE) in MAC Array}
The MAC array aims to perform the matrix multiplication $\mathbf{W}^T\mathbf{X}$ in a WS dataflow. Each PE is a 3-bit MAC unit, executing $sum_{i,j} \leftarrow x_{i,j} \times w+sum_{i-1,j}$ and $x_{i,j} \leftarrow x_{i,j-1}$. The $x_{i,j}$ and $sum_{i,j}$ are propagated rightward and downward, respectively.
In addition to MAC operation, some operations in ViT, such as quantization or biasing after the MAC stage, require post-MAC arithmetic. These can be efficiently handled by appending element-wise PE at the output of the MAC array. Since the dataflow timing is preserved, these modules can operate seamlessly without requiring additional timing-matching modules.

\begin{figure}
    \centering
    \includegraphics[width=0.8\linewidth]{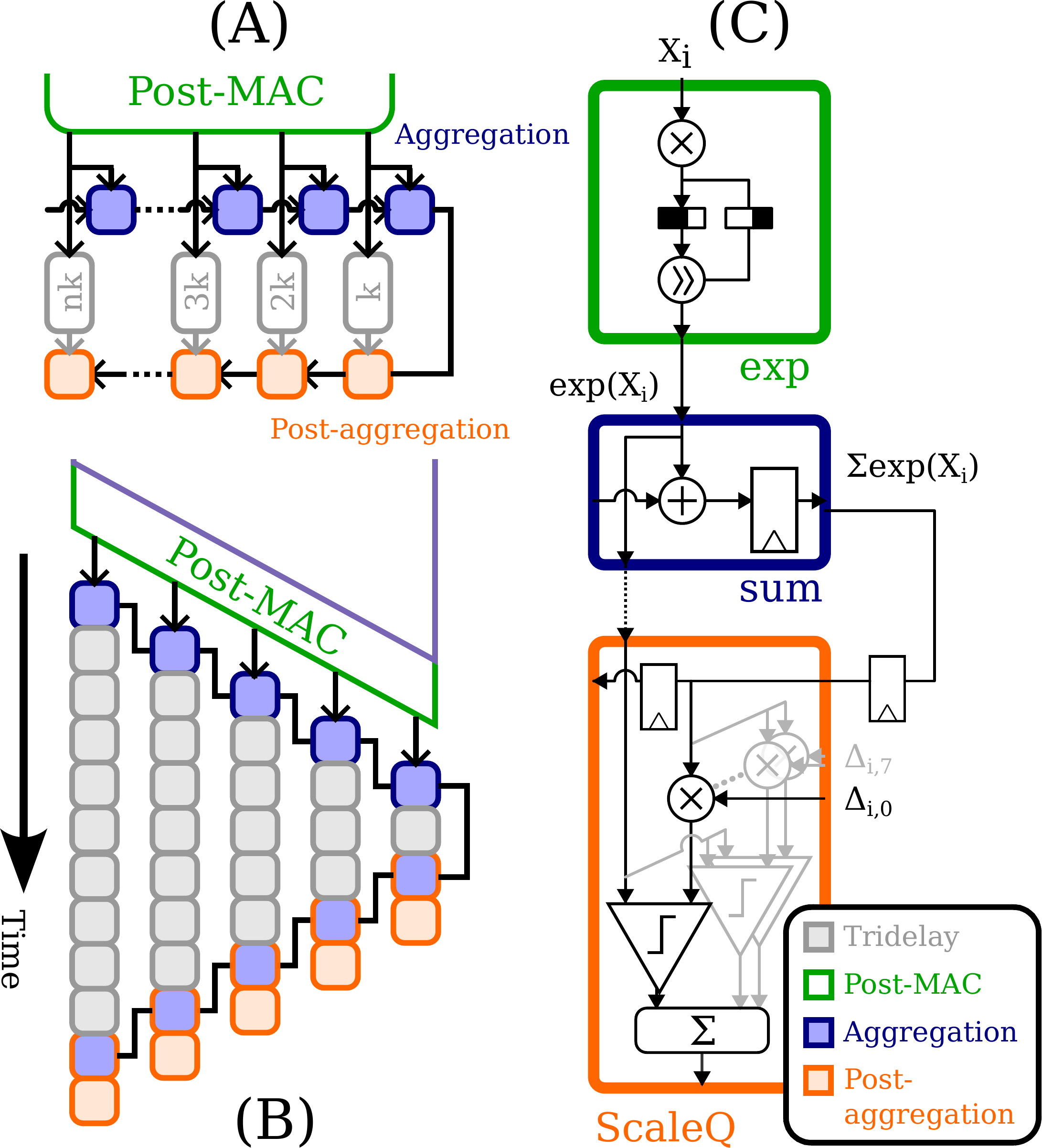}
    \caption{(A) General form of the operation (B) Corresponding timing diagram (C) A softmax quantizer can be realized with an exponential module (green), a systolic adder (blue), and a quantizer evaluates $x\underset{}{\overset{}{\gtrless}}s\times\Delta_j$, where $x = exp(\mathbf{X}_i)$ and $s = \Sigma exp(\mathbf{X}_i)$}.
    \label{fig:softmaxq}
\end{figure}

\subsubsection{1D Systolic Aggregate Operation}
Unlike convolutional neural networks, ViTs often require aggregation operation like layer normalization or softmax in MSA layers. These operations compute values across channels (e.g. mean, variance, or exponential sum). To keep high throughput, we implement these aggregation modules using a systolic-based manner.

As shown in Figure \ref{fig:softmaxq}(A), each aggregation PE combines its input with the accumulated result from previous (left) PE and passes to the next (right) PE. Once the aggregation completes, the aggregation result is propagated back to the post-aggregation module for further processing.

Timing alignment across channels is critical for systolic compatibility. Figure \ref{fig:softmaxq}(B) visualizes the timing of the aggregation: Data from each channel is delayed by one cycle relative to its previous (left) neighbor, matching the timing of the MAC array. This enable us to place one-cycle aggregation PEs in a systolic row without additional delay modules at the input.
However, the post-aggregation modules receive the aggregation result at staggered times and and therefore require timing alignment.

To address the timing alignment between aggregation and post-aggregation module, we introduce a \textit{triangular delay} module, named after its arithmetic sequence in the number of delay. This module ensures the aggregation result can be fed cycle-by-cycle in a backward order (right-to-left). An alternative design choice would be to broadcast the aggregation result in a forward order, same as the aggregation dataflow. However, this only changes the triangular delay into a parallelogram delay, which has the same cumulative delay and may further reduce the clock frequency due to longer transmission path.

Figure \ref{fig:softmaxq}(C) demonstrates an example how softmax quantization is implemented in this structure: The MAC output $\mathbf{X}_i$ is first processed by a multi-cycle exponential unit to obtain $e^{\mathbf{x}_i}$. The aggregation module then sums these exponentials and propagate the partial sums to $e^{\mathbf{X}_i}$ to compute the final $\Sigma e^{\mathbf{X}_i}$, which is propagated to all post-aggregation units again. Each post-aggregation unit then quantizes its input using a dynamic-scaled step size $\Delta_i\times\Sigma e^{\mathbf{x}_i}$ to achieve $q_{\Delta_i}(e^{\mathbf{X}_i}/\sum_i e^{\mathbf{X}_i})$. This example illustrates how complex operations can be decomposed across post-MAC, aggregation, and post-aggregation units. The details of quantizer and other modules will be discussed later in Sec. \ref{sec:op-mapping}.

The aggregation module improves the throughput by pipelining aggregation operation over multiple cycles. However, it increases the latency due to additional cycles required in inference. This additional latency cycles is proportional to the number of channels (typically ranges from 50 to 200). In our implementation, each normalization takes extra $64$ (cycle) $\times 2.5$ (ns) $=160$ (ns), which is only about $1.7\%$ of total latency. Given our throughput-centric design goals, this is an acceptable tradeoff to sustain a $400$MHz clock in our design. In contrast to latency-effected aggregation modules, post-MAC and post-aggregation modules are purely element-wise and can be executed in parallel without impacting latency, as long as the operands are available.

\begin{table}[b]
\begin{tabular}{@{}ccccc@{}}
\toprule
 & MAC & Post-MAC & Aggregation & \begin{tabular}[c]{@{}c@{}}Post-\\ aggregation\end{tabular} \\ \midrule
Q/K & WS & Scale + Bias & Normalization & NormQ \\ \midrule
V & WS & \begin{tabular}[c]{@{}c@{}}Scale + Bias\\ + Quantizer\end{tabular} & - & - \\ \midrule
A & WS + W-Load & Scale + Exp & Sum & ScaleQ \\ \midrule
Av & WS + W-Load & Quantizer & - & - \\ \bottomrule
\end{tabular}
\centering
\caption{Modules in different operations in SA}
\label{tab:operations}
\end{table}

\subsection{Operation Mapping}
\label{sec:op-mapping}
In earlier sections, we introduced the template architecture for our systolic array. Table \ref{tab:operations} summarizes the required modules within the systolic array to support various operations in a ViT model. In this section, we elabroate on how each operation is mapped to the hardware architecture.

\subsubsection{Weight-stationary Dataflow for Matrix Multiplication}

General Matrix multiplication (GeMM) acceleration has been widely studied for decades, with various architectures and dataflow proposed~\cite{luo2023calabash,chen2016eyeriss}. The choice of dataflow often depends on the target use cases and performance priorities.

\begin{figure}
    \centering
    \includegraphics[width=0.9\linewidth]{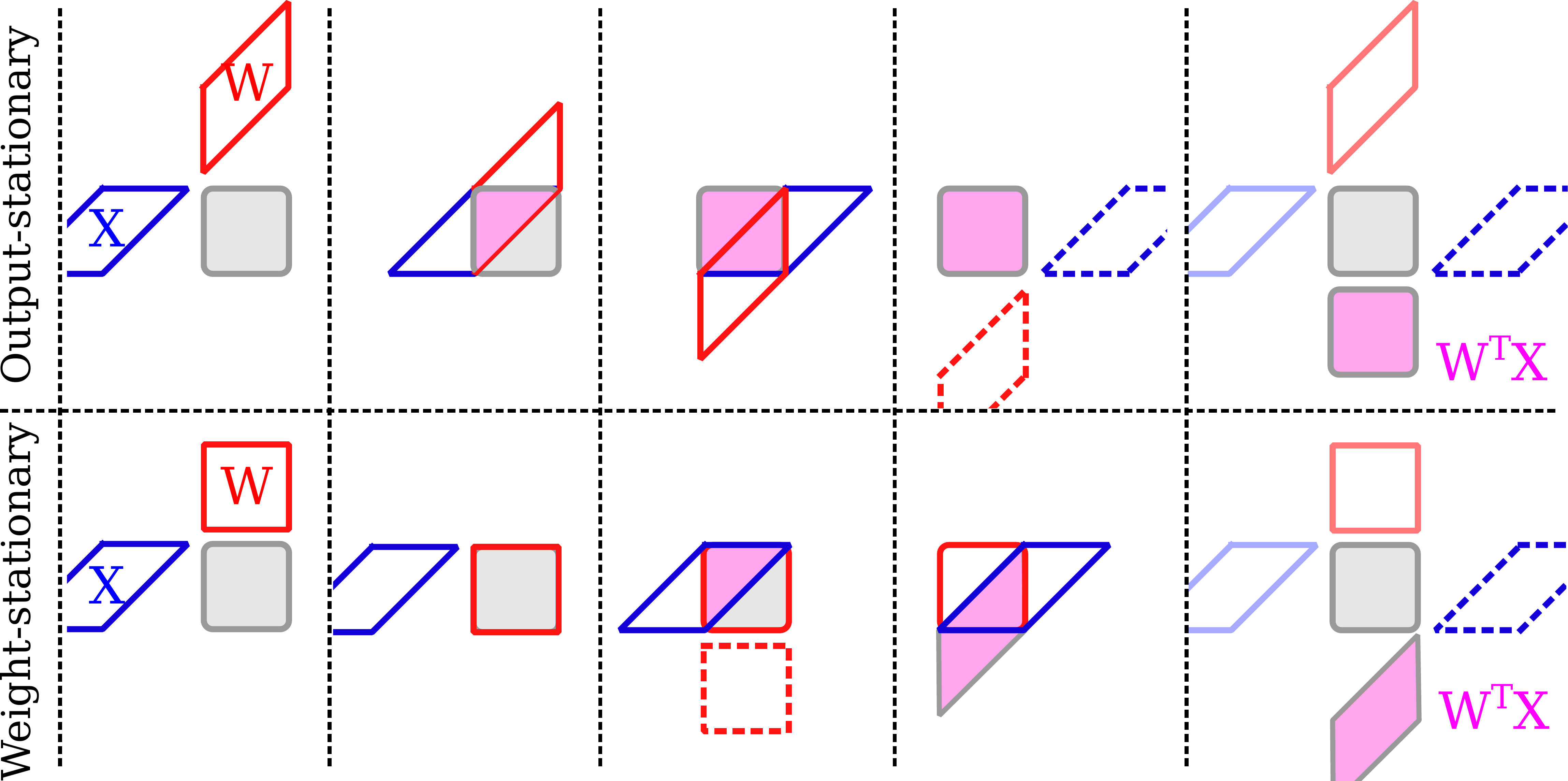}
    \caption{Comparison between output-stationary and weight-stationary dataflow. Both of them achieve the same completion time in matrix multiplication.}
    \label{fig:os-vs-ws}
\end{figure}
As this work focuses on high-throughput applications, the design only uses systolic-based weight feeding and local connection. We deliberately exclude multicast-based or broadcast-based networks in data transmission to avoid communication bottleneck. Figure \ref{fig:os-vs-ws} compares WS and output-stationary (OS) dataflow. In systolic-based architecture, both dataflow achieve the same completion time, indicating equal throughput at the system level. Our preference for WS stems from its structural similarity to the linear layers.

\begin{figure}
    \centering
    \includegraphics[width=1.0\linewidth]{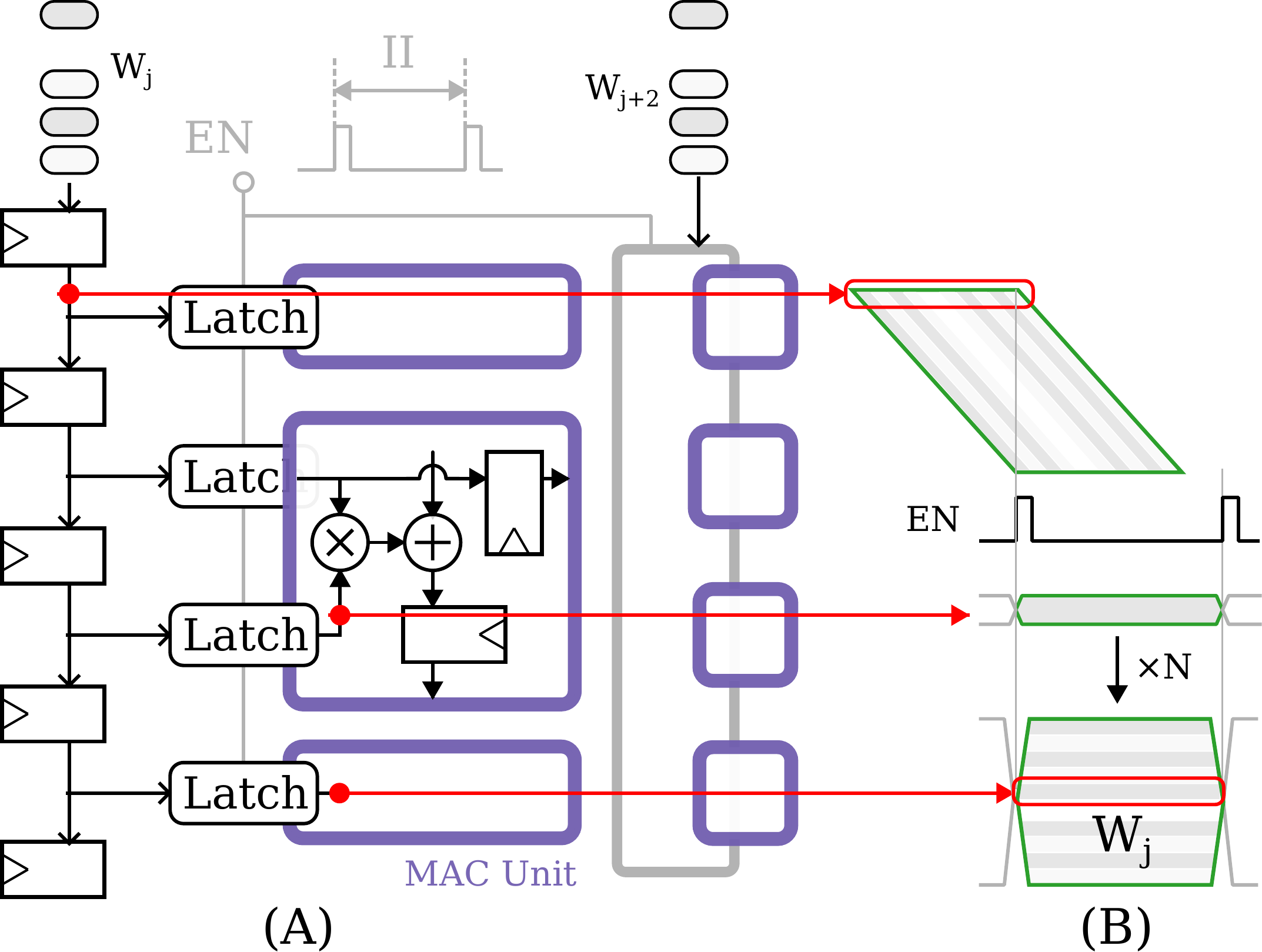}
    \caption{(A) The weight loading unit consists of a shift register chain, where each entry includes a latch that supplies weights to the corresponding MAC unit in each column. (B) In the timing diagram, each entry of the shift register forms a parallelogram. The latch enable captures a vertical slice of weights and holds it until the next enable signal, forming a rectangle whose width is equal to 1/throughput. (C) The enable signal is broadcast across the entire array, resulting in a large rectangle in the timing diagram with a height equal to the number of MACs in the array.}
    \label{fig:weight-loading}
\end{figure}

\subsubsection{Weight Loading Unit for Matrix Multiplication}

Unlike linear layers, where weights remain static and can be preloaded before the inference, dynamic matrix multiplication requires loading weights with high throughput. To meet this requirements, we introduce a dedicated weight loading unit.

As illustrated in Figure \ref{fig:weight-loading}(A), the weight loading unit consists of a shift register and a series of latches connected to intermediate registers. Each latch feeds its output into a MAC unit as the corresponding weight.

Figure \ref{fig:weight-loading}(B) shows the behavior and timing diagram of this unit. Weights are serially input into the shift register, forming a parallelogram pattern with the same (color-coded) value propagating diagonally. Once the enable signal is triggered, the values in the shift register are latched and held by the latches. This content-holding creates a rectangular pattern until the next enable is received.

\subsubsection{Quantizer Design}

The quantizer module converts high-precision data from a systolic array into low-precision values for inter-module communication. In the simplest form, it rounds the full-precision input $x$ to an integer with step size $\Delta_x$, i.e., $\lceil x/\Delta_x\rfloor$, with the $\lceil\cdot\rfloor$ a rounding function.

Given our use of 3-bit quantization, we implement $2^3-1=7$ comparators in parallel comparing the input with 7 thresholds $\Delta_{x,i} = (i +\frac{1}{2})\Delta_x$, denoted as $x \lessgtr \Delta_{x,i}$, for $i=0, 1,\dots,6$. Each boolean comparator results are combined to a 3-bit integer as the quantizer output. 

We also implement several quantizer variants for more complex cases. For example, the ScaleQ used in Figure \ref{fig:softmaxq}(C) supports $x \lessgtr s\times\Delta_x$, where $s$ is provided by another module. This enables combined softmax and quantization operations, by setting $x=exp(\mathbf{z}_i)$ and $s=\Sigma_iexp(\mathbf{z}_i)$

\subsubsection{Exponential module}
Exponential operation is widely used and has been approximated in various works~\cite{li2022q,kim2021bert,huang2023integer,li2023vit,marino2023me}. Most approaches rely on the following decomposition

\begin{equation}
    e^x = 2^{x\times log_2e} = 2^{\lfloor x\times log_2e\rfloor} \times 2^{x\times log_2e - \lfloor x\times log_2e\rfloor}
    \label{eqn:exp}
\end{equation}

The first term $2^{\lfloor x\times log_2e\rfloor}$, can be implemented efficiently with bit shift. The second term, which lies in range $(-1,0]$, is approximated in various ways, ranging from first-order \cite{li2022q}, second-order \cite{kim2021bert, huang2023integer}, to shift-based~\cite{li2023vit}. In our design, we adopt a first-order approximation, namely

\begin{equation}
    2^x \approx \frac{1}{2}x +\frac{1}{2}
    \label{eqn:exp2}
\end{equation}

As the approximation in Equation \ref{eqn:exp} is exact in $x=-1$ and $x=0$, we use it to linearly approximate the power-2 with the range $(-1,0]$. Our experiments showed that the more complex approximations did not significantly improve the accuracy.

The implementation is illustrated in Figure \ref{fig:softmaxq} (C): The prescaled ($1024$x) input is multiplied by $log_2e$. The most significant bits are used for the shift, while the remaining bits approximate $2^x$ as in Equation \ref{eqn:exp2}. The half-scaling and addition of Equation \ref{eqn:exp2} can be implemented by 1-bit right shift and single-bit overwrite. Then the approximation is passed to the shifter for the final result (Equation \ref{eqn:exp}).

It's also worth noting that the exponential results are used in softmax as $softmax(x) = e^x / \Sigma e^x$. Since this result is normalized and the scaling factors cancel out, there is no need to undo the initial prescaling.

\subsubsection{Layer Normalization + Quantization}

\begin{figure}
    \centering
    \includegraphics[width=0.9\linewidth]{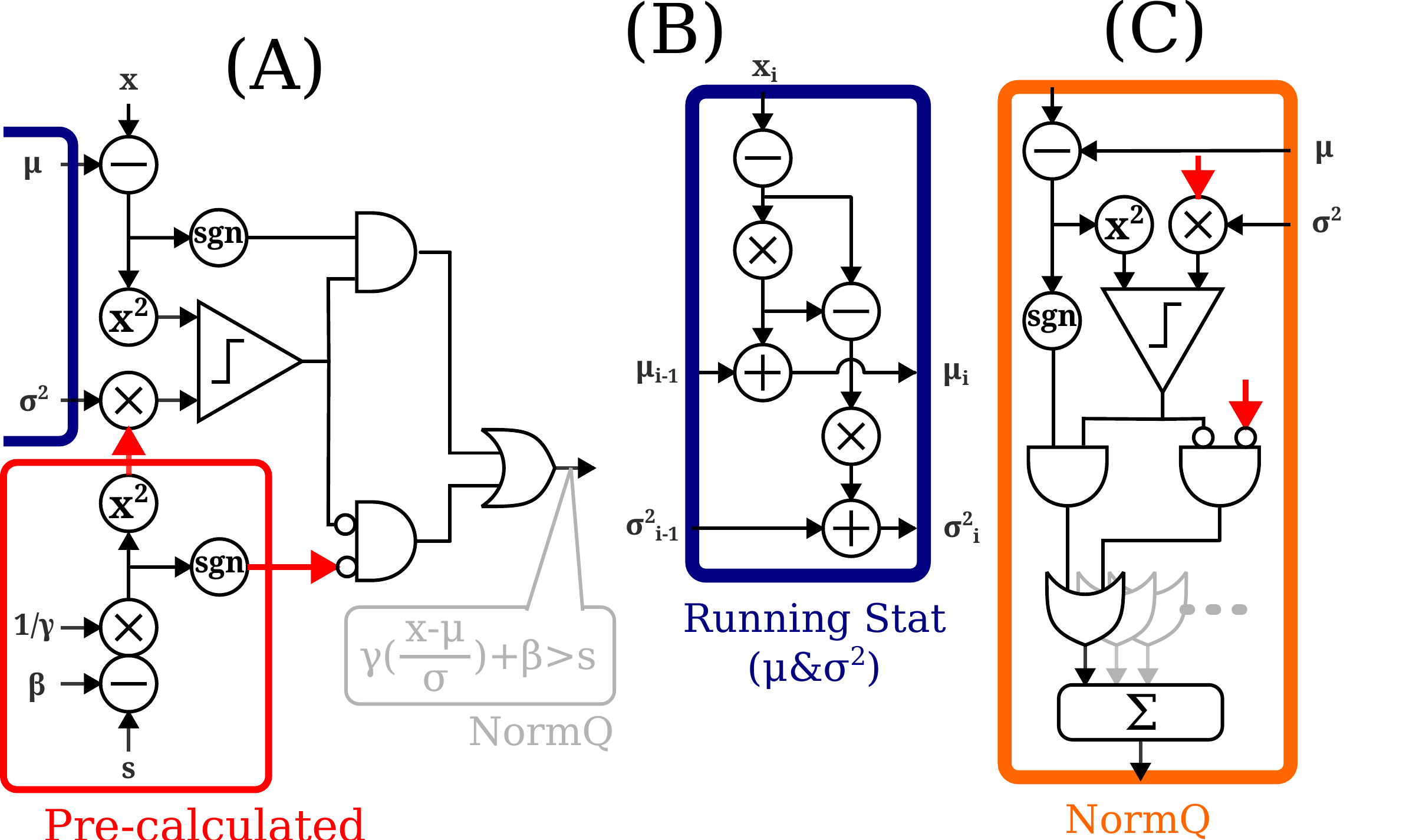}
    \caption{(A) Division-free and Square-root free normalization and quantizer (B) Systolic running statistic module (C) NormQ module to realize $\Sigma_j$ $[\gamma(x-\mu)/\sigma + \beta\underset{}{\overset{}{\gtrless}}s_j]$}
    \label{fig:norm}
\end{figure}

Layer normalization is one of the most complex operation in ViT pipeline. To compute $q_{\Delta S}(\gamma\times(x-\mu)/\sigma + \beta)$, the first challenge lies in the division and square root in variance $\sigma^2$. This can be solved by algebra tricks~\cite{lin2025low}, which results in a two-stage logic in Figure \ref{fig:norm}(A): The first stage performs the comparison 
\begin{equation}
    (x-\mu)^2\underset{}{\overset{}{\gtrless}}((s-\beta)\times(1/\gamma))^2\times\sigma^2
    \label{eqn:sqcomparison}
\end{equation}
and extracts the sign of $x-\mu$ and $(s-\beta)/\gamma$, and the second stage combines the comparator results and the sign to the correct output. We do not report the detailed derivation about this logic but this can be derived from K-map.

The second difficulty is to compute the $\mu$ and $\sigma^2$ computation in a systolic-friendly manner. We use \chingyi{Welford's algorithm \cite{welford1962note} to compute $\mu$ and $\sigma^2$ in an online manner following}
\begin{equation}
\label{eqn:running-stat}
\begin{split}
    \begin{cases}
        \mu_i = \mu_{i-1} + \frac{\mathbf{x}_i - \mu_{i-1}}{i}\\
        \sigma^2_i = \sigma^2_{i-1} + (\mathbf{x}_i-\mu_{i})(\mathbf{x}_i-\mu_{i-1})\\
    \end{cases}
\end{split}
\end{equation}
\chingyi{with initial condition $\mu_0 = 0$ and $\sigma^2_0 = 0$.} Equation \ref{eqn:running-stat} relies on current input \chingyi{$\mathbf{x}_i$} and the previous \chingyi{mean $\mu_{i-1}$ and variance $\sigma^2_{i-1}$} to calculate \chingyi{current mean $\mu_{i}$ and variance $\sigma^2_{i-1}$}, making them well-suited to a systolic implementation. The aggregation module is shown in Figure \ref{fig:norm}(B)

The output of the online statistics module is sent to the NormQ module, illustrated in Figure \ref{fig:norm}(C). NormQ module performs the division-free and square root-free quantizer based on Equation \ref{eqn:sqcomparison}, using pre-computed parameter $(s-\beta)^2/\gamma^2$ and $(s-\beta)/\gamma$. Functionally, the first stage NormQ behaves like a ScaleQ quantizer with input $(x-\mu)^2$, scale $\sigma^2$, step size $(s-\beta)^2/\gamma^2$. Two boolean sign signal $x-\mu>0$ and $(s-\beta)/\gamma>0$ are combined via logic gates to determine the final output.

\begin{figure}
    \centering
    \includegraphics[width=0.9\linewidth]{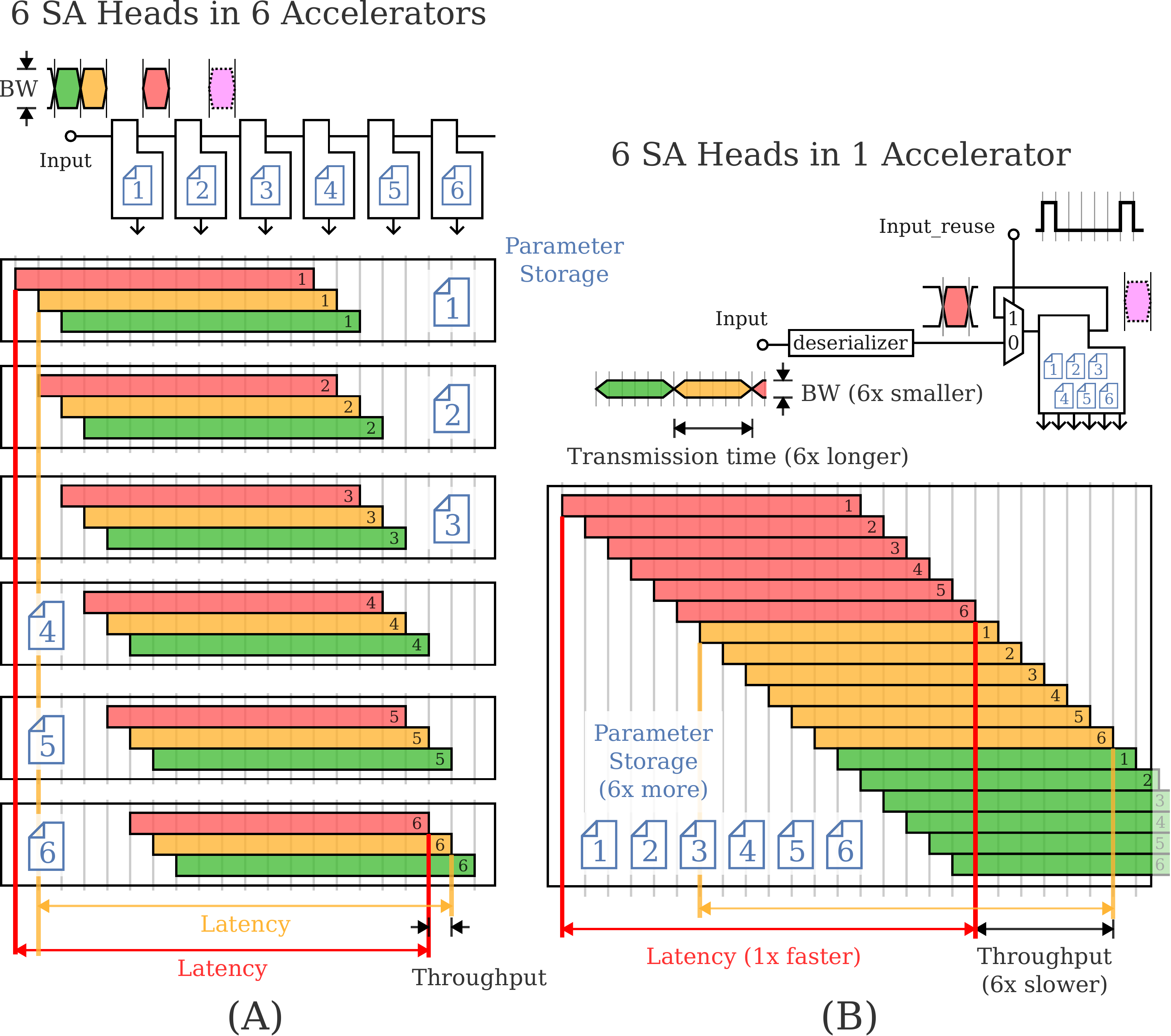}
    \caption{(A) Fully unrolled SA modules, where each SA head is mapped to dedicated hardware. (B) Time-multiplexed SA modules, where a single hardware unit reuses the same input and applies different parameters sequentially.}
    \label{fig:acc-pipelining}
\end{figure}

\subsection{Accelerator Pipelining}

The proposed architecture implements a SA module for ViT. A straightforward approach to executing MSA is to parallelize SA accelerators, with each one dedicated to processing a single SA and storing its parameter, as illustrated in Figure \ref{fig:acc-pipelining}(A). In this configuration, the accelerators can be daisy-chained to reduce communication bandwidth. However, this implementation requires a large number of resources ($>$10M LUT) and cannot fit most commercial FPGAs.

Although our design support extension to multi-FPGA due to our low communication bandwidth, we also provide accelerator pipelining to fit all SA operations into a single accelerator.

Figure \ref{fig:acc-pipelining}(B) demonstrates the pipelined accelerator idea: The input is selected either from external data source or reused from previous inputs through an input selector. The accelerator maintains the parameters for all SA heads and switches between them as needed. Pipelining accelerator does not increase the latency compared to our daisy-chain baseline. However, the throughput is reduced because the next input (yellow) has to wait until the current input (red) has been processed by all SA stages. In the case of DeiT-S ($H=6$), this results in a 6 times slower throughput. This provides a tradeoff between number of resources (or area) and throughput.

Another benefit of this pipelining technique is smaller bandwidth. As each input remains active in the accelerator for a longer time, the same data can be transmitted over a longer period. This effectively reduces the bandwidth requirement by $H$ times for the pipelining technique.

Pipelining accelerator is especially beneficial for FPGA-based system because most commercial FPGAs provide fast computational modules and plenty of DSPs, but only few FPGAs support high-speed interface such as HBM. This observation also aligns with recent trends in ASIC about the growing compute-to-communication ratio~\cite{gholami2024ai}. With operation-level pipelining, our approach provides a tradeoff between communication bandwidth, area, and throughput in inference accelerators.

\section{Results}
\subsection{Models and Training Details}
We evaluate our experiments using \chingyi{ImageNet}. Prior work has demonstrated that quantizing a pretrained model often yields higher accuracy than training from scratch~\cite{dosovitskiy2020image}. \chingyi{For ImageNet experiments, we leverage a pretrained quantized model~\cite{li2022q} on ImageNet. The difference between the quantized model and our integerized model is the per-channel and the global step size. Thus, we initialize our model with the pretrained parameter used in quantized model, and finetune for only 20 epochs with learning rate 1e-5 and Adam optimizer.}

\subsection{Experimental Setup}
To validate our hardware design, we implement the proposed system on the AMD Alveo™ U250 Data Center Accelerator FPGA, which is based on 16nm FinFET technology. We synthesize our design using Vivado 2024.2. Our implementation includes a pipelined accelerator with an on-FPGA PCIe communication module. In the following section, we report performance metrics such as timing, power, and area number from the synthesis report. While more detailed analysis methods are available, we consider the synthesis results sufficiently accurate and representative for validation purposes.

\begin{figure}
    \centering
    \includegraphics[width=0.9\linewidth]{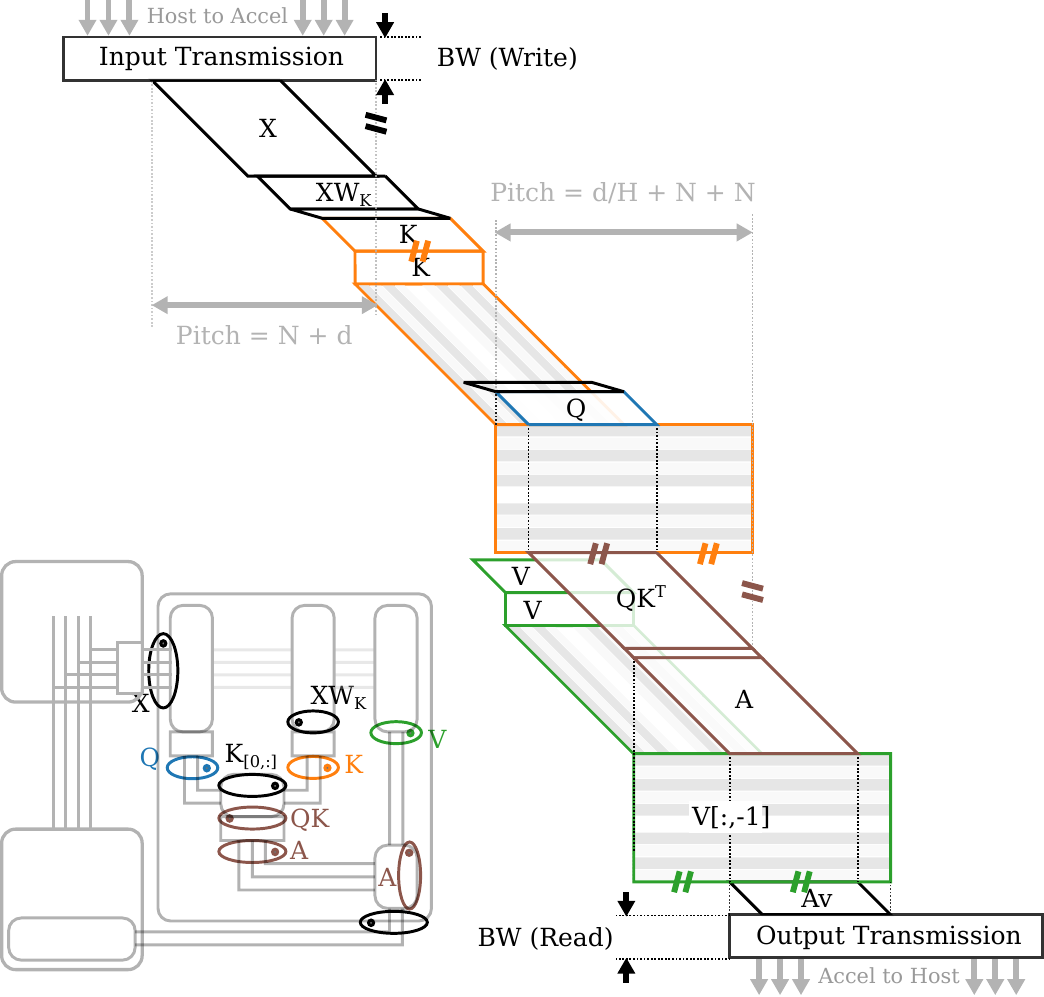}
    \caption{Timing diagram for a SA module. Its throughput depends on two factors: (1) N + d cycles from systolic input (2) d/H + N + N cycles from matmul parameter latch}
    \label{fig:timing-diagram}
\end{figure}

\subsection{Timing Performance}
\label{sec:timing-performance}
The two most important metrics in timing are latency and throughput. As our design is based on a cycle-based pipeline, both metrics can be analytically derived from the timing diagram.

Figure \ref{fig:timing-diagram} shows the timing diagram of an SA operation. After the input is transmitted, the $d$-channel $\mathbf{z}$ enters the array in a systolic dataflow, showing a parallelogram in the timing diagram. Then the post-MAC result ($\mathbf{z}\mathbf{U}_k$) and normalized low-bit key ($\mathbf{K}_{3b}$) are generated. The $\mathbf{K}_{3b}$ is later reordered from systolic order to parallel order in preparation for the first matrix multiplication ($\mathbf{QK^T}$), where it serves as the weight. This matrix $\mathbf{QK^T}$ is later passed through softmax and quantization module to produce $A$, which is finally used in the second matrix multiplication to compute the SA result $A^T\mathbf{V}$.

\subsubsection{Communication-Free case}

We begin by analyzing a simplified communication-free case with single SA unit. Shown in Figure \ref{fig:timing-diagram}, the throughput of a single SA depends on two factors: (1) the length of systolic input $\mathbf{z}$ (2) the total duration the $\mathbf{K}_{3b}$ must be held in the weight loading unit. The former spans $N+d$ cycles, while the latter must hold $\mathbf{K}_{3b}$ from the start of the $\mathbf{Q}$ input until the end of matrix multiplication $\mathbf{QK}^T$, totally $d/H + N + N$ cycles.

In DeiT-S ($N=198$, $d=384$, and $H=6$), the total interval between tokens (or "pitch") is $max(N+d, d/H+2N) = 582$ (cycles). This can be translated to $1.46\mu s$/token or $684.9$ million tokens per second, in this ideal scenario.

The latency of this communication-free, single SA operation, from the first bit of $\mathbf{z}$ to the final bit of $A\mathbf{V}$, can be computed analytically in the unit of cycles as
\begin{equation}
    d + 3\times\frac{d}{H}+\frac{d}{H}\times(MUL+1)+3\times N+5\times MUL+24
    \label{eqn:sa-latency}
\end{equation}
Here we alias $MUL$ as the number of cycles for multipliers. On the U250 FPGA, $MUL$ equals $1$, since each full-precision multiplication fits in a single DSP48E2 unit. For later DSP-free implementation, this value may increase due to multi-cycle multiplication.

Using Equation \ref{eqn:sa-latency} above, a single SA module in DeiT-S takes $1,327$ cycles. To scale this up to all heads in a MSA module, the total communication-free latency becomes
\begin{equation}
    Latency_H = Latency_1 + (H-1)\times Throughput
    \label{eqn:latency-6head}
\end{equation}

Using the number obtained above, we get the communication-free MSA latency $1327 + (6-1)\times 460 = 3627$ cycles.

\subsubsection{Communication-Aware case}
To include communication effects, we first estimate the total data transfer time. For a 64-bit/cycle bus, the transmission time for the entire MSA input is $(3\times N\times d)/64=3564$ cycles.

Accelerator pipelining allows this communication to be amortized across $H$ heads. For DeiT-S, the effective time for communication per head is $3564/6=594$ cycles. Combining with the analysis in communication-free, the actual interval, or $1/$throughput, becomes
\begin{equation}
    max(N+d, d/H+2N, (3\times N\times d)/(64\times H))
    \label{eqn:throughput}
\end{equation}
Our design, being communication-bound, has an interval of $(3\times N\times d)/(64\times H)) = 594$ cycles. Running at $400$MHz, this implies a communication bandwidth $64$ (bit/cycle) $\times 400$(MHz)$=3.125$(GB/s), and a throughput $1.49\mu s$/token or $673.4$ million tokens per second.

To compute full latency, we add the communication costs to communication-free latency which results in $1327+(6-1)\times594+3564\times2=11425$ cycles or $35.63$($\mu$s) under $400$MHz clock. This summation is valid because the communication time ($3564$) is smaller than the computation time ($1327+(6-1)\times594$), allowing communication of the current MSA to overlap with the computation of the previous MSA while avoiding pipeline bubbles.

\begin{figure}
    \centering
    \includegraphics[width=0.75\linewidth]{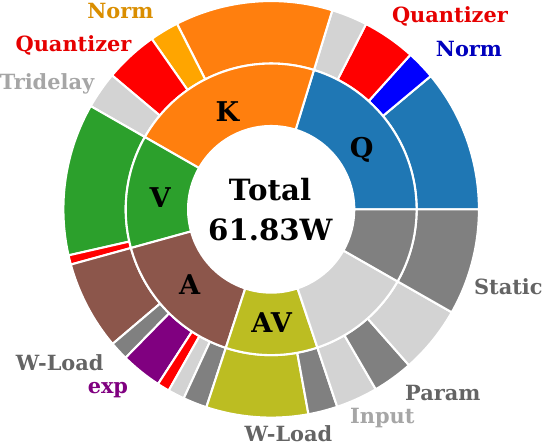}
    \caption{Power breakdown: Each section is color-coded or labeled based on their functions. The same outer-ring color means the MAC array of the inner-ring operation. For example, green outer ring is the 2D MAC array power of $\mathbf{V}$. The red and purple color represent quantizer and exponential units.}
    \label{fig:power}
\end{figure}

\begin{table}[!bht]
\begin{tabular}{@{}cccccccc@{}}
\toprule
 &  & \multicolumn{4}{c}{Module-level} & \multicolumn{2}{c}{Per-PE} \\
 &  & kLUT & kReg & DSP & Power & LUT & Power \\ \midrule
\multirow{5}{*}{Q} & MAC & 392.2 & 343.2 & 0 & 6.84 & 16 & 0.28 \\
 & Norm & 11.4 & 14.1 & 128 & 1.39 & 178 & 21.8 \\
 & NormQ & 8.6 & 17.3 & 512 & 2.56 & 134 & 40 \\
 & Others & 7.9 & 12.2 & 64 & 1.53 & - & - \\ \cmidrule(l){2-8} 
 & Subtotal & 420.1 & 386.8 & 704 & 12.3 & - & - \\ \midrule
\multirow{5}{*}{K} & MAC & 392.2 & 343.2 & 0 & 7.57 & 16 & 0.31 \\
 & Norm & 11.4 & 14.1 & 128 & 1.39 & 178 & 21.8 \\
 & NormQ & 8.6 & 17.3 & 512 & 2.56 & 134 & 40 \\
 & Others & 8.4 & 14.6 & 64 & 1.53 & - & - \\ \cmidrule(l){2-8} 
 & Subtotal & 420.5 & 389.2 & 704 & 13.1 & - & - \\ \midrule
\multirow{4}{*}{V} & MAC & 392.2 & 343.3 & 0 & 7.30 & 16 & 0.30 \\
 & Quantizer & 2.6 & 5.1 & 448 & 0.45 & 41 & 7.00 \\
 & Others & 1.1 & 3.3 & 0 & 0.00 & - & - \\ \cmidrule(l){2-8} 
 & Subtotal & 395.9 & 351.8 & 448 & 7.74 & - & - \\ \midrule
\multirow{5}{*}{A} & MAC & 190.1 & 202.7 & 0 & 5.42 & 15 & 0.43 \\
 & Softmax & 35.0 & 12.3 & 198 & 1.98 & 177 & 10 \\
 & SoftmaxQ & 17.8 & 0.0 & 0 & 0.59 & 90 & 2.99 \\
 & Others & 7.0 & 46.8 & 7 & 1.70 & - & - \\ \cmidrule(l){2-8} 
 & Subtotal & 250.0 & 261.7 & 205 & 9.70 & - & - \\ \midrule
\multirow{4}{*}{SA} & MAC & 202.8 & 320.2 & 0 & 6.30 & 16 & 0.50 \\
 & Quantizer & 3.1 & 0.0 & 448 & 0.13 & 48 & 2.00 \\
 & Others & 12.7 & 77.0 & 0 & 0.00 & - & - \\ \cmidrule(l){2-8} 
 & Subtotal & 215.5 & 397.2 & 0 & 6.43 & - & - \\ \midrule
\multicolumn{2}{c}{Others} & 17.2 & 204.5 & 0 & 12.6 & - & - \\ \midrule
\multicolumn{2}{c}{\begin{tabular}[c]{@{}c@{}}Total\\ (Utilization)\end{tabular}} & \begin{tabular}[c]{@{}c@{}}1722\\ (99.6\%)\end{tabular} & \begin{tabular}[c]{@{}c@{}}1991\\ (58\%)\end{tabular} & \begin{tabular}[c]{@{}c@{}}2509\\ (20\%)\end{tabular} & 61.8 & - & - \\ \bottomrule
& & & & & & & \\
\end{tabular}
\caption{Area and power in module-level and PE-level. The MAC operation dominates both area and power usage in MSA operation. The entire design is limited by the number of LUTs in U250 FPGA}
\label{tab:area-power-breakdown}
\end{table}

\subsection{Power analysis}
The total power consumption of our design is 61.83 (W), with a detailed breakdown shown in Figure \ref{fig:power}. This breakdown categorized the power into five primary operations ($\mathbf{Q}$, $\mathbf{K}$, $\mathbf{V}$, $A$, and $A\mathbf{V}$), along with supporting periphery (input selection/delay, parameter feeding, and others) and static device power.

The breakdown shows that the 2D MAC array power accounts for the majority of the power consumption across all primary operations. Among these, the least dominant one is the $A$. This is primarily due to its complicated post-MAC operations using shifter-based exponential computations, and relatively large portion of post-MAC PEs; specifically one post-MAC PE every $d/H=64$ 3b-MAC PEs.

In contrast to the power dominance of the MAC array, quantization and other post-MAC operations contribute only a small portion to the overall power. Although these units consume more power per PE, their total contribution remain limited due to the lower number of PEs. This observation validates the idea of our design strategy, which concentrates optimization efforts on the compute-intensive modules.

Beyond the module-level view, we also perform a detailed analysis at the PE-level. Table \ref{tab:area-power-breakdown} reports the per-PE power consumption for various modules. The 3-b MAC PE in linear layers ($\mathbf{Q}$, $\mathbf{K}$, and $\mathbf{V}$) consume $0.28-0.31$W per PE, while PE in matrix multiplication consumes more power, ranging from $0.43-0.50$W per PE. The increased power is attributed to the additional weight loading unit in matrix multiplication array.

\subsection{Area analysis}

To demonstrate the area efficiency of our integer-based design, we also report hardware resource usage. Table \ref{tab:area-power-breakdown} presents the number of look-up tables (LUTs), registers, and DSPs used across different modules. At the first glance, it is evident that our design is limited by LUT availability. Then, similar to the power analysis, the MAC array also dominates in terms of area utilization. In fact, the MAC array’s dominance in area is even more pronounced than its dominance in power.

We also examine area at the PE-level to estimate the theoretical performance. The 3-b MAC PE, whether used in either linear layers or matrix multiplication, takes between $15$ to $16$ LUTs per PE. The slight discrepancy between the actual and projected ($16\times$ number of PE) module-level LUT count arises from reduced partial sum bitwidth in the first few rows of the MAC array. Additionally, we find the extra weight loading unit in PEs of matrix multiplication do not use any LUT except for the global enable.

The average $15.87$ LUTs per MAC PE sets an upper bound for the scalability on a single FPGA. Given a total $1,728,000$ LUTs on U250, the maximum number of 3-b MAC PE can be derived from $1,728,000 / 15.87 = 108,885$. In our design, we implement $3\times384\times64+2\times64\times198=99,072$ PEs, which occupies about $90.9\%$ of the total LUT resources.

\begin{figure}
    \centering
    \includegraphics[width=0.9\linewidth]{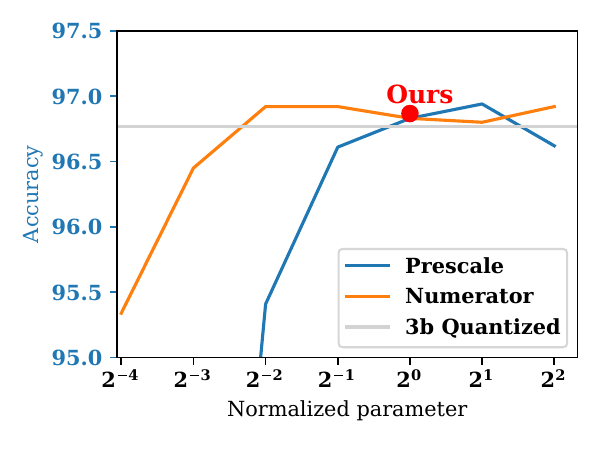}
    \caption{The accuracy vs parameters. Lower prescale and numerator results in quantization error and further reduces accuracy. Our parameter choice achieves the model accuracy closed quantized 3-bit ViT}
    \label{fig:param-sweep}
\end{figure}

\subsection{Error Analysis}
Since the accuracy of 3-bit quantized DeiT-S has been reported compared to full-precision DeiT-S~\cite{li2022q}, our analysis focuses on the accuracy gap between the 3-bit quantized DeiT-S~\cite{li2022q} and our 3-bit integerized DeiT-S.

Most of our integerization process is algebraic equivalent to the quantized model, which means the results should be identical under the assumption of infinite precision. The only deviation is the approximation of some step sizes. Specifically, we use a channel-shared $\overline{\Delta_X}$ rather than channel-wise ${\Delta_X}$. From our experiments, this approximation introduces no observable loss in accuracy.

In our actual implementation, the primary source of error in our design originates from the precision error from normalization. In order to fit each multiplication into the $27\times18$ multiplier of the on-board DSP48E2 unit, we apply scaling in normalization. Derived from Equation \ref{eqn:running-stat}, our fixed-point normalization has the form

\begin{equation}
\label{eqn:running-stat-fixed-point}
\begin{split}
    \begin{cases}
        \mu_i = \mu_{i-1} + [(2^\nu/i)(s\times x_i - \mu_{i-1})]/2^\nu\\
        \sigma^2_i = \sigma^2_{i-1} + (s\times x_i-\mu_{i-1})(s\times x_i-\mu_{i})
    \end{cases}
\end{split}
\end{equation}

In Equation \ref{eqn:running-stat-fixed-point}, two additional terms are introduced: $2^\nu$, referred to as the \textit{numerator}, and $s$, referred to as the \textit{prescale factor}. The \textit{numerator} increases the precision of the fractional division by scaling $1/i$ up ($2^\nu$x) before the multiplication and scaling down afterward. The scaling-up and scaling-down, achieved from shifter, are lossless, but rounding $2^\nu / i$ to an integer introduces quantization error in inference.

The \textit{prescale factor} $s$ is applied to input $x$ before normalization. This increases $\mu$ by a factor of $s$ and $\sigma^2$ by a factor of $s^2$, requiring a scale-back afterwards too. Similar to $2^\nu$, the subsequent scale-back is also achieved with right shift.

In our implementation, we set $2^\nu=64$ and $s=32$. To evaluate the impact of these parameters by sweeping a range of values normalized to our chosen baseline. As shown in Figure \ref{fig:param-sweep}, the accuracy of our design remain close to that of the previously reported quantized model~\cite{li2022q}, confirming that our integerization introduce no observable errors. On the other side, when either prescale factor $s$ or numerator $2^\nu$ drop below a certain threshold, we observe accuracy degradation. This loss comes from the quantization error from $\lceil 2^\nu/i\rfloor$ and $\lceil s\times x_i - \mu_{i-1}\rfloor$ respectively.

\subsection{Accuracy Comparison}

\chingyi{To validate the accuracy-power tradeoff of our integerized model, we evaluate the accuracy of DeiT series model~\cite{touvron2021training} and power of the 3-bit integerized design.}

\chingyi{To estimate power across hardware using different bitwidth, we adopt the widely used quadratic relationship between multiplier power and bitwidth~\cite{jaiswal2015low,sarangpure2017design}. Combined with the total number of OPs required per model, the estimated power can be normalized as
\begin{equation}
\text{Power}\propto\text{OPs}\times(\text{bitwidth}/8)^2
\end{equation}
}
\begin{figure}
    \centering
    \includegraphics[width=0.9\linewidth]{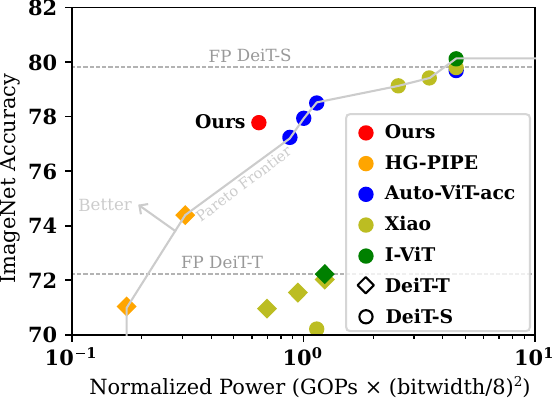}
    \caption{\chingyi{Accuracy Comparison: We compare our 3-bit integerized model with HG-PIPE~\cite{guo2024hg}, Auto-ViT-acc~\cite{li2022auto}, Xiao~\cite{xiao2025refining}, and I-ViT. Our model achieves lower normalized power at the similar accuracy level, pushing the Pareto frontier out in accuracy-power plot.}}
    \label{fig:acc}
\end{figure}

\chingyi{Figure \ref{fig:acc} compares the accuracy and normalized power of our model against several prior works, including HG-PIPE~\cite{guo2024hg}, Auto-ViT-acc~\cite{li2022auto}, Xiao~\cite{xiao2025refining}, and I-ViT~\cite{li2023vit}. We exclude quantization works that rely on dequantized operands, as their computation does not benefit from low-bit arithmetic.}

\chingyi{In Figure \ref{fig:acc}, accuracy and normalized power forms a Pareto frontier from the bottom-left to top-right. The closest point in accuracy to our model is a configuration from Auto-ViT-acc~\cite{li2022auto}, which uses a mix of 3-bit and 4-bit weight along with 4-bit activations. When applying only 3-bit weight in their design, the accuracy drops below ours while still more power consuming. This push in the Pareto frontier demonstrates the effectiveness of our integerization algorithm in achieving superior power efficiency without sacrificing accuracy.}

\begin{table*}[bht]
\caption{Comparison to other works: We compare our design to other similar works in the ascending order of power efficiency. The derived numbers are labeled with "*". Due to the lighter operation (INT3), we achieve the fastest clock and the smallest bandwidth. The systolic array also gives us dense computation, resulting in the high power and the high compute intensity. We also achieve the best power efficiency (GOPs/s/W) among all the transformer accelerator.}
\centering
\begin{tabular}{cccccccccccc}
\hline
 & Board & Tech & Model & \begin{tabular}[c]{@{}c@{}}OP\\ Type\end{tabular} & \begin{tabular}[c]{@{}c@{}}clk\\ (MHz)\end{tabular} & \begin{tabular}[c]{@{}c@{}}Power\\ (W)\end{tabular} & \begin{tabular}[c]{@{}c@{}}Peak\\ GOPs/s\end{tabular} & \begin{tabular}[c]{@{}c@{}}Actual\\ GOPs/s\end{tabular} & \begin{tabular}[c]{@{}c@{}}GOPs/s/W\\ (GOPs/J)\end{tabular} & \begin{tabular}[c]{@{}c@{}}Bandwidth\\ (Read/Write)\end{tabular} & OP/Byte \\ \hline
ViA~\cite{wang2022via} & U50 & 16nm & Swin-T & FP16 & 300 & 39 & - & 309.6 & *7.9 & - & - \\ \hline
DFX~\cite{hong2022dfx} & U280 & 16nm & 345M (GPT-2) & FP16 & - & 45 & - & 184.1 & *4.1 & 400 & *0.46 \\ \hline
Fan~\cite{fan2022adaptable} & VCU128 & 16nm & FABNet & FP16 & 200 & 11.4 & *128.0 & *128.0 & *11.3 & *$\sim$50 & *1.28 \\ \hline
Ye~\cite{ye2023accelerating} & U250 & 16nm & Rush~\cite{rush2018annotated} & INT8 & 300 & 77.2 & 4915 & 1800 & *23.3 & 12 / 9 & *150 \\ \hline
\multirow{2}{*}{Huang\cite{huang2023integer}} & \multirow{2}{*}{ZCU102} & \multirow{2}{*}{16nm} & ViT-T & \multirow{2}{*}{INT8} & 300 & - & 1229 & 616.1 & - & - & - \\
 &  &  & ViT-S &  & 300 & 29.6 & 1229 & 762.7 & *25.8 & - & - \\ \hline
\multirow{2}{*}{HeatViT~\cite{dong2023heatvit}} & \multirow{2}{*}{ZCU102} & \multirow{2}{*}{16nm} & DeiT-S &  & - & 10.7 & - & *221 & 20.6 & - & - \\
 &  &  & DeiT-B &  & - & 11.4 & - & *425 & 37.4 & - & - \\ \hline
Lu~\cite{lu2020hardware} & VU13P & 16nm & \begin{tabular}[c]{@{}c@{}}Transformer\\ (MHA)\end{tabular} & INT8 & 200 & 16.7 & 819 & *629 & 37.7 & - & - \\ \hline
Ftrans~\cite{li2020ftrans} & VCU118 & 16nm & \begin{tabular}[c]{@{}c@{}}RoBERTa\\base\end{tabular} & INT16 & - & 25.1 & - & *1077 & *43.0 & 7.88 & 33.49 \\ \hline
\multirow{2}{*}{Zhang~\cite{zhang2024109}} & \multirow{2}{*}{ZCU9EG} & \multirow{2}{*}{16nm} & ViT-S & \multirow{2}{*}{FP8} & \multirow{2}{*}{300} & *21.4 & - & *1150 & *53.7 & - & - \\
 &  &  & ViT-B &  &  & *21.4 & - & *1254 & *58.6 & - & - \\ \hline
\multirow{2}{*}{ME-ViT~\cite{marino2023me}} & \multirow{2}{*}{U200} & \multirow{2}{*}{16nm} & \multirow{2}{*}{ViT-B} & \multirow{2}{*}{INT8} & 150 & 17.8 & - & *1149 & *64.5 & *5.74 & 200 \\
 &  &  &  &  & 300 & 31.8 & - & *2298 & *72.2 & *11.49 & 200 \\ \hline
\multirow{2}{*}{Calabash~\cite{luo2023calabash}} & VU9P & \multirow{2}{*}{16nm} & \multirow{2}{*}{BERT} & \multirow{2}{*}{INT16} & 243 & 12.9 & *995 & 880 & *68.2 & *30.38 & 28.97 \\
 & ZCU102 &  &  &  & 300 & \textbf{6.6} & *614 & 530 & *80.3 & *37.5 & 14.13 \\ \hline
\multirow{3}{*}{Xiao~\cite{xiao2025refining}} & \multirow{3}{*}{U250} & \multirow{3}{*}{16nm} & Deit-T & \multirow{3}{*}{\begin{tabular}[c]{@{}c@{}}Custom\\ INT5\end{tabular}} & 183 & 40.1 &  & 1488 & *37.1 & - & - \\
 &  &  & DeiT-S &  & 192 & 28.4 &  & 2861 & *100.8 & - & - \\
 &  &  & DeiT-B &  & 179 & 25.3 &  & 2332 & *92.4 & - & - \\ \hline
\multirow{4}{*}{Auto-ViT-acc\cite{li2022auto}} & \multirow{4}{*}{ZCU102} & \multirow{4}{*}{16nm} & DeiT-S & \multirow{2}{*}{\begin{tabular}[c]{@{}c@{}}FP3.5\\ +FP4\end{tabular}} & \multirow{4}{*}{150} & 10.3 & - & *711 & *68.8 & - & - \\
 &  &  & DeiT-B &  &  & 11.0 & - & *996 & *90.3 & - & - \\ \cline{5-5}
 &  &  & DeiT-S & \multirow{2}{*}{FP3.5} &  & 6.6 & - & *689 & *105.2 & - & - \\
 &  &  & DeiT-B &  &  & 8.1 & - & *989 & *122.0 & - & - \\ \hline
\multirow{4}{*}{HG-PIPE~\cite{guo2024hg}} & ZCU102 & \multirow{4}{*}{16nm} & \multirow{3}{*}{DeiT-T} & \multirow{2}{*}{FP4} & 375 & 21.9 & - & 1974 & *90.1 & 8.53 & *231 \\ \cline{2-2}
 & \multirow{3}{*}{VCK190} &  &  &  & \textbf{425} & 43.4 & - & 4536 & *104.5 & 8.53 & *532 \\ \cline{5-5}
 &  &  &  & \multirow{2}{*}{FP3} & \textbf{425} & 46.7 & - & 8898 & *190.5 & 8.53 & *1043 \\ \cline{4-4}
 &  &  & DeiT-S &  & 350 & 48.1 & - & 6854 & *142.5 & 8.53 & *803 \\ \hline
Ours & U250 & 16nm & \begin{tabular}[c]{@{}c@{}}3b DeiT-S\\ (MHA)\end{tabular} & INT3 & 400 & 61.83 & \textbf{42515} & \textbf{13568} & \textbf{219.4} & \textbf{3.13 / 3.13} & \textbf{4342} \\ \hline
\end{tabular}
\label{tab:results}
\end{table*}

\subsection{Compare to Same-technology Accelerators}
Table \ref{tab:results} compares our accelerator to other FPGA-based transformer implementations in terms of clock rate, power, gigaoperations per second (GOPs/s), power efficient (GOPs/s/W), bandwidth (GB/s), and operational intensity (OP/byte). \chingyi{Here we use GOPs/s instead of tokens/s to allow cross-model comparison.}

Due to the reduced logic used in 3-bit MAC, our design can accommodate more operations within the same hardware resources. Combined with a $400$ MHz operating frequency, we achieve the highest throughput $13,568$ (GOPs/s). This result is even comparable to GPUs using the same technology (Tesla P100: $18,700$ GOPs/s in FP16 and GeForce GTX 1080: $8,873$ GOPs/s in FP32). While comparing low-bit integer operations on FPGAs with FP operations on GPUs is not entirely fair, these results demonstrate the potential of low-bit models.

In terms of power efficiency, our design exceeds that of models using 8-bit or higher precision by more than 2.5x. Even when compared to low-bit accelerator~\cite{guo2024hg}, we still have better power efficiency and throughput together. This improvement is driven both by the simplicity of integer operations and our model integerization approach.

Finally, we also optimize bandwidth to improve performance within the roofline model~\cite{williams2009roofline}. The combination of low-bit packet and accelerator pipelining enables our design to operate with just $3.13$ (GB/s) of bandwidth, a level that is achievable through PCIe 3.0 x4 protocol. At the same time, our design also delivers an operational intensity of $4342$ (OP/Byte), which is $28$x higher than the existing state-of-the-art works, further explaining the superior performance in the roofline model.

\subsection{Compare with Other Technology Nodes}

\chingyi{We also compare our design against prior works implemented in different technologies. We follow fixed-frequency, constant-field Dennard scaling~\cite{dennard2003design} for a fair comparison. The following relationships are used for normalization
\begin{equation}
    \begin{split}
    \begin{cases}
        \text{Throughput}\propto f\times\text{Density}\times\text{Die size}\\
        \text{Power efficiency}\propto \text{GOPs}/\text{J}\approx 1/(\text{C}_g\text{V}^2)
    \end{cases}
    \end{split}
\end{equation}
According to Dennard scaling, when scaling to a more advanced technology by a factor of $\alpha$ ($\alpha > 1$), transistor density, gate capacitance $C_g$, and supply voltage $V$ scale as $\alpha^2$, $\alpha$, and $1/\alpha$ respectively. This  estimation results in theoretical improvement of $\alpha^2\times$ in throughput and $\alpha\times$ in power efficiency.
}

\begin{figure}[bt]
    \centering
    \includegraphics[width=\linewidth]{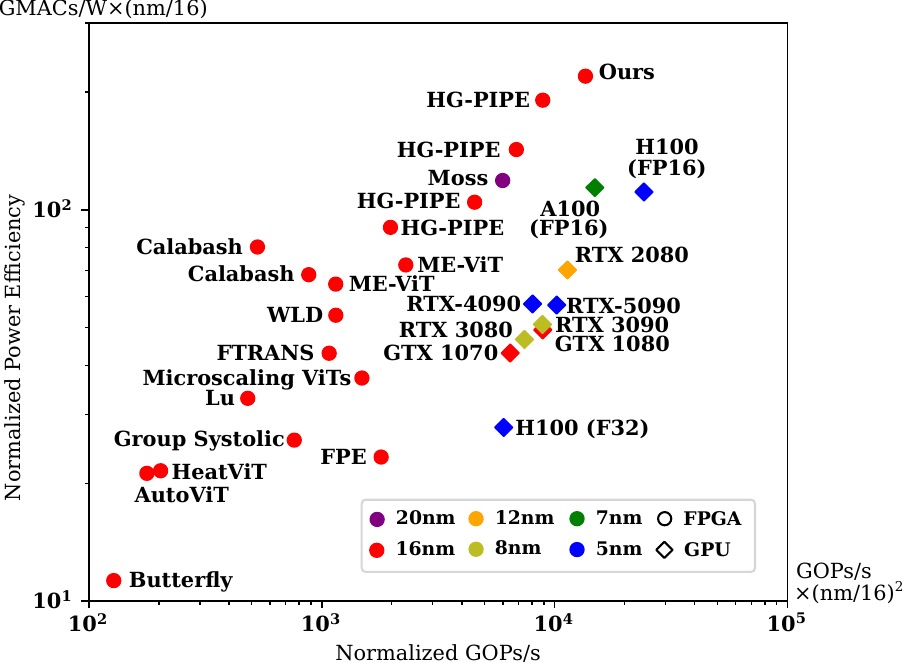}
    \caption{Comparison of normalized power efficiency and throughput. Our design achieves the highest normalized power efficiency among all evaluated works and ranks just below NVIDIA's H100 and A100 GPUs in terms of normalized throughput}
    \label{fig:comp-cross-tech}
\end{figure}

\chingyi{
Figure \ref{fig:comp-cross-tech} presents a comparison of power efficiency and throughput across various FPGA-based transformer accelerators and GPUs from different generation. The GPU specifications are obtained from TechPowerUp database~\cite{TechPowerUp}. In throughput, some GPUs have higher throughput than our design. This mainly comes from die size-dependent throughput as die size is not reported in FPGAs. In contrast, our design achieves the highest normalized power efficiency among all evaluated works, validating the effectiveness of our low-bit architecture.
}

\begin{table}[]
\caption{Latency estimation for the full DeiT-S, where N, H, d, and M are number of words, number of heads, embedding dimension, and MLP scaling, respectively}
\begin{tabular}{@{}cccccc@{}}
\toprule
 & & Formula & DeiT-T & DeiT-S & DeiT-B \\ \midrule 
\multicolumn{2}{c}{Comm} & $3\times N\times d/(64\times H)$ & 1782 & 3564 & 7128 \\ \midrule
{\parbox[t]{1mm}{\multirow{3}{*}{\rotatebox[origin=c]{90}{MSA}}}} & SA (6x) & Eqn \ref{eqn:latency-6head} & 1729 & 4297 & 22963 \\
 & Comm & $3\times N\times d/(64\times H)$ & 1782 & 3564 & 7128 \\
 & Projection & $2\times d+N$ & 582 & 966 & 1734 \\ \midrule
\multicolumn{2}{c}{Comm} & $3\times N\times d/(64\times H)$ & 1782 & 3564 & 7128 \\ \midrule
\multicolumn{2}{c}{MLP} & $(M+2)\times d+N$ & 1350 & 2502 & 4806 \\ \midrule
\multicolumn{2}{c}{Comm} & $3\times N\times d/(64\times H)$ & 1782 & 3564 & 7128 \\ \midrule
\multicolumn{2}{c}{Layer (1x)} & \begin{tabular}[c]{@{}c@{}}$4\times comm+$\\ $MSA+MLP$\end{tabular} & 10789 & 22021 & 58015\\ \midrule
{\parbox[t]{1mm}{\multirow{2}{*}{\rotatebox[origin=c]{90}{Total}}}} & Cycles (k) & $12\times$Layer & 129.5 & 264.3 & 696.2 \\
 & Latency ($\mu$s) & 12$\times$Layer$\times$2.5 & 323.7 & 660.6 & 1740.5 \\
\bottomrule
\label{tab:deit-full}
\end{tabular}
\end{table}

\subsection{Extended to Full DeiT-S and Other Models}
Section \ref{sec:timing-performance} discussed the timing performance of the MSA in DeiT-S. In this discussion, we want to provide an estimation to the entire model to show the potential of our low-bit integer inference architecture.

To make a conservative estimation, we assume the MLP operation is implemented in another accelerator. Thus, the MHA and MLP results are transformed back and forth to MLP and MHA accelerators with the costs of four transmission. 

Table \ref{tab:deit-full} lists the latency estimation for the full DeiT-S, as well as two of its variant DeiT-T and DeiT-B.

\begin{table}[]
\caption{Latency comparison between different works}
\begin{tabular}{@{}ccccccc@{}}
\toprule
 & \multicolumn{3}{c}{Latency ($\mu$s)} & \multicolumn{3}{c}{GOPs} \\
 & DeiT-T & DeiT-S & DeiT-B & DeiT-T & DeiT-S & DeiT-B \\ \midrule
Ours (INT3) & \textbf{324} & \textbf{661} & \textbf{1741} & 1.23 & 4.57 & 17.54 \\
\cite{dong2023heatvit} (FP8) & 3687 & 9158 & 18248 & \textbf{0.75} & \textbf{2.02} & \textbf{7.75} \\
\cite{xiao2025refining} (FP5) & 1697 & 3234 & 15138 & 1.23 & 4.57 & 17.54 \\ \bottomrule
\label{tab:deit-latency}
\end{tabular}
\end{table}

\begin{figure}[t]
    \centering
    \includegraphics[width=0.9\linewidth]{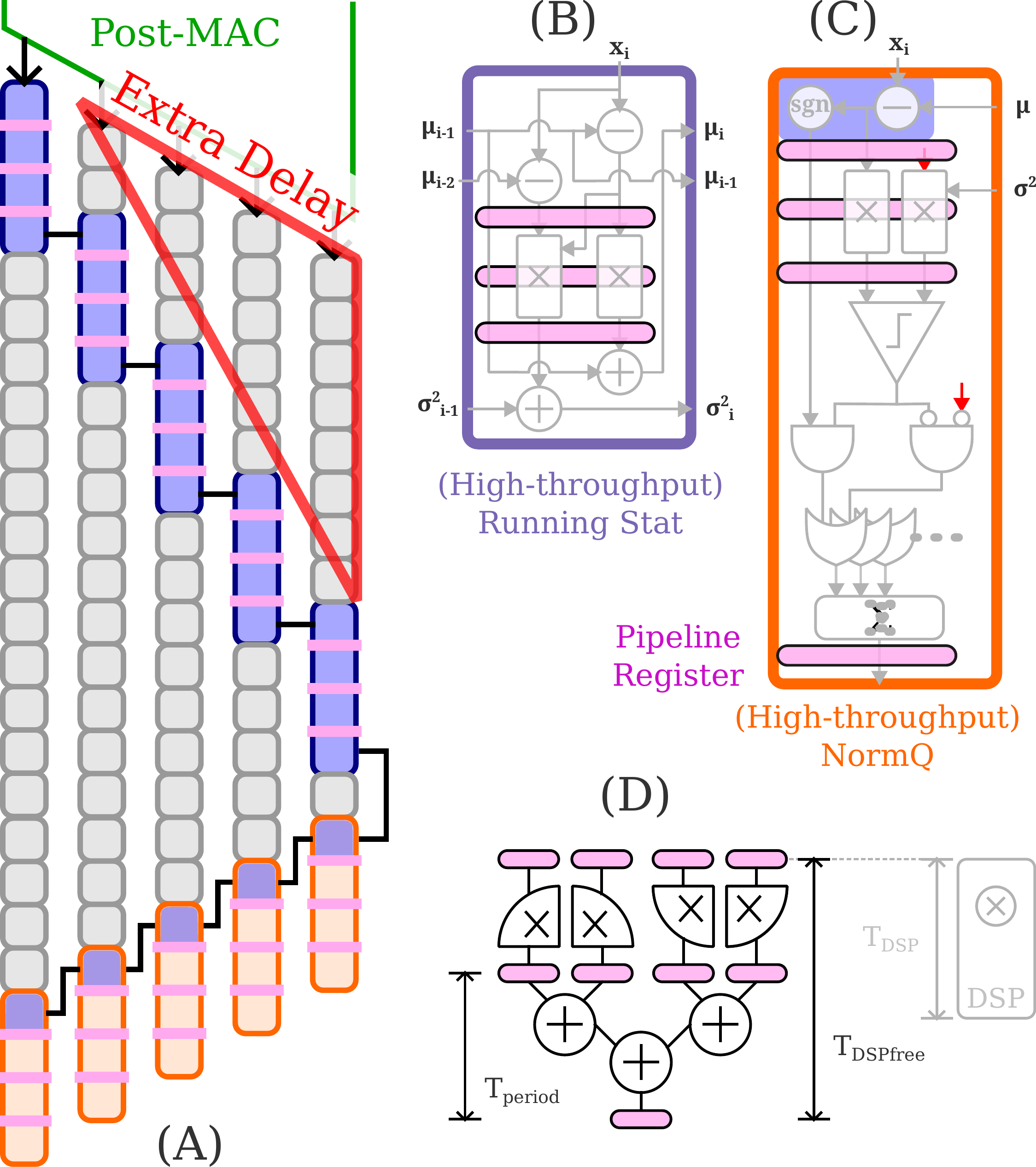}
    \caption{(A) High-throughput normalization and quantization: The pipelined aggregation uses more but shorter cycles to achieve higher throughput. This technique also incurs additional latency and requires extra delay elements for timing alignment. (B)(C) High-throughput running statistics and quantization unit: The multiplier is pipelined to shorten the clock period. (D) The multiplier pipeline consists of two stages: the first stage performs partial products in parallel, and the second stage accumulates them. While the total latency ($T_{DSPfree}$) exceeds a single DSP delay ($T_{DSP}$), the pipeline enables a shorter clock period $T_{period}$ than $T_{DSP}$}
    \label{fig:dsp-free}
\end{figure}

\begin{table}[]
\caption{Comparison between DSP-involved and DSP-free design}
\begin{tabular}{@{}ccccccc@{}}
\toprule
 & \begin{tabular}[c]{@{}c@{}}Freq\\ (MHz)\end{tabular} & \begin{tabular}[c]{@{}c@{}}Power\\ (W)\end{tabular} & \begin{tabular}[c]{@{}c@{}}BW\\ (GB/s)\end{tabular} & \begin{tabular}[c]{@{}c@{}}Latency\\ ($\mu$s)\end{tabular} & TOPs/s & GOPs/W \\ \midrule
DSP & 400 & \textbf{61.83} & \textbf{3.13} & 28.56 & 13.57 & \textbf{219.5} \\
DSP-free & 400 & 96.91 & \textbf{3.13} & 28.74 & 13.57 & 140.0 \\
DSP-free & \textbf{500} & 125.5 & 3.91 & \textbf{22.99} & \textbf{17.37} & 128.4 \\ \bottomrule
\end{tabular}
\label{tab:dsp-free}
\end{table}

\subsection{Higher-throughput DSP-free Design}
\label{sec:dsp-free}
One of the primary computational bottleneck in our design is the $400$MHz clock, which is limited by the timing constraints of the DSP units. To address this limitation, we also explore an alternative DSP-free design to achieve higher clock frequencies.

While removing DSPs allows for faster clocking, it introduces a tradeoff for increased latency due to a deeper pipeline using more cycles. Figure \ref{fig:dsp-free} (A) shows the timing diagram for the DSP-free aggregation. As the aggregation unit takes additional cycles, it requires an extra delay cycle to match the timing between the post-MAC unit and the aggregation stage.

Figure \ref{fig:dsp-free} (B)(C) illustrates the DSP-free statistics and NormQ. The multiplier is divided into multiple cycles, and the rest of them remains unchanged. Figure \ref{fig:dsp-free} (D) visually demonstrates the tradeoff for divided multipliers. Although the total computation time in the DSP-free version $T_{DSPfree}$ is longer than in the original DSP-based version $T_{DSP}$, the resulting clock period $T_{period}$ is shorter than $T_{DSP}$. This enables a faster clock and thus higher throughput overall.

Table \ref{tab:dsp-free} compares the performance between the DSP-free and DSP-based design. The DSP-free design achieves a higher clock frequency and increased power consumption. To isolate the impact of clock frequency, we also evaluate the DSP-free design under the same $400$ MHz clock. Across both frequency settings, the DSP-free designs exhibits similar power efficiency and GOP per cycle. The version with a $500$ MHz clock achieves proportionally higher power and throughput (GOPs/s) with the faster clock.

At the same frequency, the DSP-free design consumes more power. This is because of the lower energy efficiency of LUT-based arithmetic compared to DSP units. In terms of latency, the multipliers in DSP-free design requires between $2-4$ cycles depending on operand size. This introduces additional $1\times64 + 4\times1+1\times3=71$ cycles to complete a single SA.

Importantly, the above additional latency does not scale with the number of heads $H$. In the context of the full MSA pipeline, the overhead cost is only $71/11425=0.6\%$, which is small and negligible.

Another drawback of the DSP-free design is its reduced power efficiency. Replacing DSP with look-up table and pipeline register takes extra 56.7\% power cost. In server-based application, this increased power is equivalent to increased cost from the same inference, limiting the practicality of adopting DSP-free architectures despite their frequency advantages.


\section{Conclusions}

\chingyi{In this paper, we presented a low-bit, systolic array-based architecture to enhance both throughput and power efficiency for transformer model inference. The proposed architecture employs systolic-compatible units to construct a deep and fine-grained pipeline. The operational intensity is optimized by low communication-overhead transmission point and large compute density from simple low-bit arithmetic. We validate our approach by implementing the system on an FPGA (Alveo U250). The system achieves 77.81\% top-1 accuracy on ImageNet, delivers a throughput of 13,568 GOPs/s at 61.83 W. The power efficiency 219.4 GOPs/s/W outperforms state-of-the-art solutions and demonstrates the effectiveness of our low-bit, systolic-based design for efficient transformer inference.}

\bibliographystyle{IEEEtran}
\bibliography{references}

\begin{thebibliography}{10}
\providecommand{\url}[1]{#1}
\csname url@samestyle\endcsname
\providecommand{\newblock}{\relax}
\providecommand{\bibinfo}[2]{#2}
\providecommand{\BIBentrySTDinterwordspacing}{\spaceskip=0pt\relax}
\providecommand{\BIBentryALTinterwordstretchfactor}{4}
\providecommand{\BIBentryALTinterwordspacing}{\spaceskip=\fontdimen2\font plus
\BIBentryALTinterwordstretchfactor\fontdimen3\font minus \fontdimen4\font\relax}
\providecommand{\BIBforeignlanguage}[2]{{%
\expandafter\ifx\csname l@#1\endcsname\relax
\typeout{** WARNING: IEEEtran.bst: No hyphenation pattern has been}%
\typeout{** loaded for the language `#1'. Using the pattern for}%
\typeout{** the default language instead.}%
\else
\language=\csname l@#1\endcsname
\fi
#2}}
\providecommand{\BIBdecl}{\relax}
\BIBdecl

\bibitem{Todorovic.2024}
\BIBentryALTinterwordspacing
I.~Todorovic, ``Chatgpt consumes enough power in one year to charge over three million electric cars,'' \emph{Balkan Green Energy News}. [Online]. Available: \url{https://balkangreenenergynews.com/chatgpt-consumes-enough-power-in-one-year-to-charge-over-three-million-electric-cars/}
\BIBentrySTDinterwordspacing

\bibitem{Moss.2024}
\BIBentryALTinterwordspacing
S.~Moss, ``Openai training and inference costs could reach $7bn for 2024, ai startup set to lose $5bn - report,'' \emph{Data Center Dynamices}. [Online]. Available: \url{https://www.datacenterdynamics.com/en/news/openai-training-and-inference-costs-could-reach-7bn-for-2024-ai-startup-set-to-lose-5bn-report/}
\BIBentrySTDinterwordspacing

\bibitem{samsi2023words}
S.~Samsi, D.~Zhao, J.~McDonald, B.~Li, A.~Michaleas, M.~Jones, W.~Bergeron, J.~Kepner, D.~Tiwari, and V.~Gadepally, ``From words to watts: Benchmarking the energy costs of large language model inference,'' in \emph{2023 IEEE High Performance Extreme Computing Conference (HPEC)}.\hskip 1em plus 0.5em minus 0.4em\relax IEEE, 2023, pp. 1--9.

\bibitem{sardana2023beyond}
N.~Sardana, J.~Portes, S.~Doubov, and J.~Frankle, ``Beyond chinchilla-optimal: Accounting for inference in language model scaling laws,'' \emph{arXiv preprint arXiv:2401.00448}, 2023.

\bibitem{jacob2018quantization}
B.~Jacob, S.~Kligys, B.~Chen, M.~Zhu, M.~Tang, A.~Howard, H.~Adam, and D.~Kalenichenko, ``Quantization and training of neural networks for efficient integer-arithmetic-only inference,'' in \emph{Proceedings of the IEEE conference on computer vision and pattern recognition}, 2018, pp. 2704--2713.

\bibitem{lin2025low}
C.-Y. Lin and S.~Shah, ``Low-bit integerization of vision transformers using operand reodering for efficient hardware,'' \emph{arXiv preprint arXiv:2504.18547}, 2025.

\bibitem{li2023vit}
Z.~Li and Q.~Gu, ``I-vit: Integer-only quantization for efficient vision transformer inference,'' in \emph{Proceedings of the IEEE/CVF International Conference on Computer Vision}, 2023, pp. 17\,065--17\,075.

\bibitem{li2022q}
Z.~Li, T.~Yang, P.~Wang, and J.~Cheng, ``Q-vit: Fully differentiable quantization for vision transformer,'' \emph{arXiv preprint arXiv:2201.07703}, 2022.

\bibitem{williams2009roofline}
S.~Williams, A.~Waterman, and D.~Patterson, ``Roofline: an insightful visual performance model for multicore architectures,'' \emph{Communications of the ACM}, vol.~52, no.~4, pp. 65--76, 2009.

\bibitem{luo2017thinet}
J.-H. Luo, J.~Wu, and W.~Lin, ``Thinet: A filter level pruning method for deep neural network compression,'' in \emph{Proceedings of the IEEE international conference on computer vision}, 2017, pp. 5058--5066.

\bibitem{han2015deep}
S.~Han, H.~Mao, and W.~J. Dally, ``Deep compression: Compressing deep neural networks with pruning, trained quantization and huffman coding,'' \emph{arXiv preprint arXiv:1510.00149}, 2015.

\bibitem{courbariaux2016binarized}
M.~Courbariaux, I.~Hubara, D.~Soudry, R.~El-Yaniv, and Y.~Bengio, ``Binarized neural networks: Training deep neural networks with weights and activations constrained to+ 1 or-1,'' \emph{arXiv preprint arXiv:1602.02830}, 2016.

\bibitem{rastegari2016xnor}
M.~Rastegari, V.~Ordonez, J.~Redmon, and A.~Farhadi, ``Xnor-net: Imagenet classification using binary convolutional neural networks,'' in \emph{European conference on computer vision}.\hskip 1em plus 0.5em minus 0.4em\relax Springer, 2016, pp. 525--542.

\bibitem{alemdar2017ternary}
H.~Alemdar, V.~Leroy, A.~Prost-Boucle, and F.~P{\'e}trot, ``Ternary neural networks for resource-efficient ai applications,'' in \emph{2017 international joint conference on neural networks (IJCNN)}.\hskip 1em plus 0.5em minus 0.4em\relax IEEE, 2017, pp. 2547--2554.

\bibitem{zhou2016dorefa}
S.~Zhou, Y.~Wu, Z.~Ni, X.~Zhou, H.~Wen, and Y.~Zou, ``Dorefa-net: Training low bitwidth convolutional neural networks with low bitwidth gradients,'' \emph{arXiv preprint arXiv:1606.06160}, 2016.

\bibitem{choi2018bridging}
J.~Choi, P.~I.-J. Chuang, Z.~Wang, S.~Venkataramani, V.~Srinivasan, and K.~Gopalakrishnan, ``Bridging the accuracy gap for 2-bit quantized neural networks (qnn),'' \emph{arXiv preprint arXiv:1807.06964}, 2018.

\bibitem{lin2021fq}
Y.~Lin, T.~Zhang, P.~Sun, Z.~Li, and S.~Zhou, ``Fq-vit: Post-training quantization for fully quantized vision transformer,'' \emph{arXiv preprint arXiv:2111.13824}, 2021.

\bibitem{kim2021bert}
S.~Kim, A.~Gholami, Z.~Yao, M.~W. Mahoney, and K.~Keutzer, ``I-bert: Integer-only bert quantization,'' in \emph{International conference on machine learning}.\hskip 1em plus 0.5em minus 0.4em\relax PMLR, 2021, pp. 5506--5518.

\bibitem{huang2023integer}
M.~Huang, J.~Luo, C.~Ding, Z.~Wei, S.~Huang, and H.~Yu, ``An integer-only and group-vector systolic accelerator for efficiently mapping vision transformer on edge,'' \emph{IEEE Transactions on Circuits and Systems I: Regular Papers}, vol.~70, no.~12, pp. 5289--5301, 2023.

\bibitem{ye2023accelerating}
W.~Ye, X.~Zhou, J.~Zhou, C.~Chen, and K.~Li, ``Accelerating attention mechanism on fpgas based on efficient reconfigurable systolic array,'' \emph{ACM Transactions on Embedded Computing Systems}, vol.~22, no.~6, pp. 1--22, 2023.

\bibitem{luo2023calabash}
Z.~Luo, L.~Lu, Y.~Jin, L.~Jia, and Y.~Liang, ``Calabash: Accelerating attention using a systolic array chain on fpgas,'' in \emph{2023 33rd International Conference on Field-Programmable Logic and Applications (FPL)}.\hskip 1em plus 0.5em minus 0.4em\relax IEEE, 2023, pp. 242--247.

\bibitem{zhang2024109}
Y.~Zhang, L.~Feng, H.~Shan, and Z.~Zhu, ``A 109-gops/w fpga-based vision transformer accelerator with weight-loop dataflow featuring data reusing and resource saving,'' \emph{IEEE Transactions on Circuits and Systems for Video Technology}, 2024.

\bibitem{li2022auto}
Z.~Li, M.~Sun, A.~Lu, H.~Ma, G.~Yuan, Y.~Xie, H.~Tang, Y.~Li, M.~Leeser, Z.~Wang, X.~Lin, and F.~Zhenman, ``Auto-vit-acc: An fpga-aware automatic acceleration framework for vision transformer with mixed-scheme quantization,'' in \emph{2022 32nd International Conference on Field-Programmable Logic and Applications (FPL)}.\hskip 1em plus 0.5em minus 0.4em\relax IEEE, 2022, pp. 109--116.

\bibitem{marino2023me}
K.~Marino, P.~Zhang, and V.~K. Prasanna, ``Me-vit: A single-load memory-efficient fpga accelerator for vision transformers,'' in \emph{2023 IEEE 30th International Conference on High Performance Computing, Data, and Analytics (HiPC)}.\hskip 1em plus 0.5em minus 0.4em\relax IEEE, 2023, pp. 213--223.

\bibitem{chen2016eyeriss}
Y.-H. Chen, T.~Krishna, J.~S. Emer, and V.~Sze, ``Eyeriss: An energy-efficient reconfigurable accelerator for deep convolutional neural networks,'' \emph{IEEE journal of solid-state circuits}, vol.~52, no.~1, pp. 127--138, 2016.

\bibitem{dosovitskiy2020image}
A.~Dosovitskiy, L.~Beyer, A.~Kolesnikov, D.~Weissenborn, X.~Zhai, T.~Unterthiner, M.~Dehghani, M.~Minderer, G.~Heigold, S.~Gelly, J.~Uszkoreit, and H.~Neil, ``An image is worth 16x16 words: Transformers for image recognition at scale,'' \emph{arXiv preprint arXiv:2010.11929}, 2020.

\bibitem{touvron2021training}
H.~Touvron, M.~Cord, M.~Douze, F.~Massa, A.~Sablayrolles, and H.~J{\'e}gou, ``Training data-efficient image transformers \& distillation through attention,'' in \emph{International conference on machine learning}.\hskip 1em plus 0.5em minus 0.4em\relax PMLR, 2021, pp. 10\,347--10\,357.

\bibitem{luo2016finegrained}
\BIBentryALTinterwordspacing
Y.~Huang, C.~Li, M.~Li, L.~Van~der Perre, and W.~Dehaene, ``Fine-grained hardware switching scheme for power reduction in multiplication,'' \emph{Electronics Letters}, vol.~52, no.~16, pp. 1374--1375, 2016. [Online]. Available: \url{https://ietresearch.onlinelibrary.wiley.com/doi/abs/10.1049/el.2015.3828}
\BIBentrySTDinterwordspacing

\bibitem{welford1962note}
B.~P. Welford, ``Note on a method for calculating corrected sums of squares and products,'' \emph{Technometrics}, vol.~4, no.~3, pp. 419--420, 1962.

\bibitem{gholami2024ai}
A.~Gholami, Z.~Yao, S.~Kim, C.~Hooper, M.~W. Mahoney, and K.~Keutzer, ``Ai and memory wall,'' \emph{IEEE Micro}, vol.~44, no.~3, pp. 33--39, 2024.

\bibitem{jaiswal2015low}
K.~B. Jaiswal, N.~Kumar, P.~Seshadri, and L.~G, ``Low power wallace tree multiplier using modified full adder,'' in \emph{2015 3rd international conference on signal processing, communication and networking (ICSCN)}.\hskip 1em plus 0.5em minus 0.4em\relax IEEE, 2015, pp. 1--4.

\bibitem{sarangpure2017design}
P.~R. Sarangpure, D.~S. Chaudhari, and Y.~D. Kapse, ``Design and implementation of wallace compressor multiplier using vedic mathematics,'' 2017.

\bibitem{guo2024hg}
Q.~Guo, J.~Wan, S.~Xu, M.~Li, and Y.~Wang, ``Hg-pipe: Vision transformer acceleration with hybrid-grained pipeline,'' in \emph{Proceedings of the 43rd IEEE/ACM International Conference on Computer-Aided Design}, 2024, pp. 1--9.

\bibitem{xiao2025refining}
C.~Xiao, J.~Cheng, and A.~Zhao, ``Refining datapath for microscaling vits,'' \emph{arXiv preprint arXiv:2505.22194}, 2025.

\bibitem{wang2022via}
T.~Wang, L.~Gong, C.~Wang, Y.~Yang, Y.~Gao, X.~Zhou, and H.~Chen, ``Via: A novel vision-transformer accelerator based on fpga,'' \emph{IEEE Transactions on Computer-Aided Design of Integrated Circuits and Systems}, vol.~41, no.~11, pp. 4088--4099, 2022.

\bibitem{hong2022dfx}
S.~Hong, S.~Moon, J.~Kim, S.~Lee, M.~Kim, D.~Lee, and J.-Y. Kim, ``Dfx: A low-latency multi-fpga appliance for accelerating transformer-based text generation,'' in \emph{2022 55th IEEE/ACM International Symposium on Microarchitecture (MICRO)}.\hskip 1em plus 0.5em minus 0.4em\relax IEEE, 2022, pp. 616--630.

\bibitem{fan2022adaptable}
H.~Fan, T.~Chau, S.~I. Venieris, R.~Lee, A.~Kouris, W.~Luk, N.~D. Lane, and M.~S. Abdelfattah, ``Adaptable butterfly accelerator for attention-based nns via hardware and algorithm co-design,'' in \emph{2022 55th IEEE/ACM International Symposium on Microarchitecture (MICRO)}.\hskip 1em plus 0.5em minus 0.4em\relax IEEE, 2022, pp. 599--615.

\bibitem{rush2018annotated}
A.~M. Rush, ``The annotated transformer,'' in \emph{Proceedings of workshop for NLP open source software (NLP-OSS)}, 2018, pp. 52--60.

\bibitem{dong2023heatvit}
P.~Dong, M.~Sun, A.~Lu, Y.~Xie, K.~Liu, Z.~Kong, X.~Meng, Z.~Li, X.~Lin, Z.~Fang, and Y.~Wang, ``Heatvit: Hardware-efficient adaptive token pruning for vision transformers,'' in \emph{2023 IEEE International Symposium on High-Performance Computer Architecture (HPCA)}.\hskip 1em plus 0.5em minus 0.4em\relax IEEE, 2023, pp. 442--455.

\bibitem{lu2020hardware}
S.~Lu, M.~Wang, S.~Liang, J.~Lin, and Z.~Wang, ``Hardware accelerator for multi-head attention and position-wise feed-forward in the transformer,'' in \emph{2020 IEEE 33rd International System-on-Chip Conference (SOCC)}.\hskip 1em plus 0.5em minus 0.4em\relax IEEE, 2020, pp. 84--89.

\bibitem{li2020ftrans}
B.~Li, S.~Pandey, H.~Fang, Y.~Lyv, J.~Li, J.~Chen, M.~Xie, L.~Wan, H.~Liu, and C.~Ding, ``Ftrans: energy-efficient acceleration of transformers using fpga,'' in \emph{Proceedings of the ACM/IEEE International Symposium on Low Power Electronics and Design}, 2020, pp. 175--180.

\bibitem{dennard2003design}
R.~H. Dennard, F.~H. Gaensslen, H.-N. Yu, V.~L. Rideout, E.~Bassous, and A.~R. LeBlanc, ``Design of ion-implanted mosfet's with very small physical dimensions,'' \emph{IEEE Journal of solid-state circuits}, vol.~9, no.~5, pp. 256--268, 2003.

\bibitem{TechPowerUp}
``Techpowerup,'' \url{https://www.techpowerup.com/}.

\end{thebibliography}

\end{document}